\pdfoutput=1

\documentclass[reprint,amsmath,amssymb,floatfix,aps,longbibliography]{revtex4-1}
\usepackage{graphics,graphicx,float}

\usepackage[pdftex,
            pdfauthor={Steven A. Frank},
            pdftitle={},
            pdfsubject={Natural selection. VII. History and interpretation of kin selection theory}]{hyperref} 

\def\citeyear{\citep}
\def\autocite{\citep}

\newcommand{\Ga}{\alpha}

\newcommand{\Gr}{\rho}

\newcommand{\Gt}{\tau}

\newcommand{\Eq}[1]{eqn~\ref{eq:#1}}

\newcommand{\boldrule}{\hrule height 1.2pt}
\newcommand{\noterule}{\medskip\boldrule\medskip}	

\newcommand{\noterulenote}[1]{\null}

\newcount\BoxNum \BoxNum 1\relax
\makeatletter
\newcommand{\boxlabel}[1]{%
  \protected@write \@auxout {}{\string \newlabel {box:#1}{{\the\BoxNum}{\thepage}{\noexpand\relax}%
  	{\@ifundefined{hyper@@anchor}{\relax}{box.\the\BoxNum}}%
  	{}}}%
  \@ifundefined{hyper@@anchor}{}{\hypertarget{box.\the\BoxNum}{}}%
  \advance\BoxNum 1\relax}
\makeatother
\newcommand{\Boxx}[1]{Box~\ref{box:#1}}
\newcommand{\BoxLabel}{Box~\the\BoxNum}

\begin{document}

\title{Natural selection. VII. History and interpretation of kin selection theory}

\author{Steven A.\ Frank}
\email[email: ]{safrank@uci.edu}
\homepage[homepage: ]{http://stevefrank.org}
\affiliation{Department of Ecology and Evolutionary Biology, University of California, Irvine, CA 92697--2525  USA}

\begin{abstract}

Kin selection theory is a kind of causal analysis.  The initial form of kin selection ascribed cause to costs, benefits, and genetic relatedness.  The theory then slowly developed a deeper and more sophisticated approach to partitioning the causes of social evolution.  Controversy followed because causal analysis inevitably attracts opposing views.  It is always possible to separate total effects into different component causes.  Alternative causal schemes emphasize different aspects of a problem, reflecting the distinct goals, interests, and biases of different perspectives.  For example, group selection is a particular causal scheme with certain advantages and significant limitations. Ultimately, to use kin selection theory to analyze natural patterns and to understand the history of debates over different approaches, one must follow the underlying history of causal analysis.  This article describes the history of kin selection theory, with emphasis on how the causal perspective improved through the study of key patterns of natural history, such as dispersal and sex ratio, and through a unified approach to demographic and social processes.  Independent historical developments in the multivariate analysis of quantitative traits merged with the causal analysis of social evolution by kin selection\footnote{\href{http://dx.doi.org/10.1111/jeb.12131}{doi:\ 10.1111/jeb.12131} in \textit{J. Evolutionary Biology}}\footnote{Reprint at \href{http://stevefrank.org/reprints-pdf/13NS07.pdf}{http://stevefrank.org/reprints-pdf/13NS07.pdf}}\footnote{Part of the Topics in Natural Selection series. See \Boxx{preface}.}.

\end{abstract}

\maketitle

\begin{quote}
As is often the case, once a topic has become in vogue, its name ceases to have meaning $\ldots$ \autocite{lazar12de-meaning}.
\end{quote}

\section*{Introduction}

Ideas are embedded in their history and language.  Hamilton's \citeyear{hamilton70selfish} theories of inclusive fitness and kin selection are good examples.  As understanding deepened, the original ideas transformed into broader concepts of selection and evolutionary process.  With that generalization, the initial language that remains associated with the topic has become distorted. The confused language and haphazard use of incorrect historical context have led to significant misunderstanding and meaningless argument.

Current understanding transcends the initial interpretation of `kin selection,' which attaches to some notion of similarity by descent from a recent common ancestor.  The other candidate phrases, such as `inclusive fitness' or `group selection,' also have problems.  We are left with a topic that derives from those antecedent notions and clearly has useful application to those biological puzzles.  At the same time, the modern understanding of altruism connects to the analysis of selection on multiple characters, to interactions between different species, and to the broadest generalizations of the theory of natural selection.  

A good scholarly history of kin selection and its descendants has yet to be written.  Here, I give a rough historical outline in the form of a nonmathematical narrative (see \Boxx{scope}).  I describe the history from my personal perspective.  Because I worked actively on the subject over several decades, I perceive the history by the ways in which my own understanding changed over time.  \Boxx{literature} highlights other perspectives and key citations.

\section*{Early history}

According to \textcite{darwin59on-the-origin}, natural selection favors traits that enhance individual reproduction. Puzzles arise when traits reduce individual reproduction while providing aid to others.  In this context, \textcite{fisher30the-genetical} discussed the problem of warning coloration in mimicry:
\begin{quote}
[D]istastefulness $\ldots$ is obviously capable of giving protection to the species as a whole, through its effect upon the instinctive or acquired responses of predators, yet since any individual tasted would seem almost bound to perish, it is difficult to perceive how individual increments of the distasteful quality, beyond the average level of the species, could confer any individual advantage.
\end{quote}
Fisher then gave one possible solution:
\begin{quote}
[W]ith gregarious larvae the effect will certainly be to give the increased protection especially to one particular group of larvae, probably brothers and sisters of the individual attacked. The selective potency of the avoidance of brothers will of course be only half as great as if the individual itself were protected; against this is to be set the fact that it applies to the whole of a possibly numerous brood. $\ldots$ The ideal of heroism has been developed among such peoples considerably beyond the optimum of personal advantage, and its evolution is only to be explained, in terms of known causes, by the advantage which it confers, by repute and prestige, upon the kindred of the hero.
\end{quote}
\textcite{haldane55population} restated Fisher's argument. \textcite{williams57natural} presented a specific and limited analysis, which they applied to altruism in social insects.  Some rather vague models considered more diffuse forms of genetic similarity created by population structuring of groups \autocite{haldane32the-causes,wright45tempo}.  These analyses hinted at a more general concept. However, none of them expressed the deeper principle in a clear and convincing way. 

\subsection*{Hamilton's rule}

\textcite{hamilton63the-evolution,hamilton64the-genetical,hamilton64the-geneticalb} initiated modern approaches by expressing a rule for the increase in an altruistic behavior
\begin{equation}\label{eq:hr}
  rB - C > 0.
\end{equation}
Here, $C$ is the cost to an actor for performing the altruistic behavior, $B$ is the benefit gained by a recipient of the altruistic act, and $r$ measures the relatedness between actor and recipient.  The idea is that the actor pays a personal cost in reduced reproduction for helping another individual, the recipient gains increased reproduction from the altruistic act, and the relatedness translates the recipient's enhanced reproduction back to the actor's valuation of that extra reproduction.  When the benefit back to the actor, $rB$, is greater than the actor's direct cost, $C$, then the behavior is favored by natural selection.  

Hamilton figured out how to measure $r$ by analyzing population genetic models.  With the methods Hamilton used in those early papers, he was only able to give a rough description for the proper measure for relatedness.  His initial expression in terms of the genetic correlation between actor and recipient was in the right direction and was later refined by \textcite{hamilton70selfish}.  

Hamilton also defined a new and more general notion of fitness, which he called `inclusive fitness.'  Instead of counting the number of offspring by an individual, inclusive fitness makes a more extensive calculation of how phenotypes influence the transmission of genes from one generation to the next.  In the inclusive fitness interpretation, \Eq{hr} describes the changes in direct and indirect reproduction associated with each change in behavior.  Thus, the inclusive fitness of a particular behavioral act is the indirect reproductive gain through the recipient, $B$, multiplied by the relatedness, $r$, minus the loss in direct reproduction, $C$.  The relatedness, $r$, measures the genetic discount of substituting the reproduction of the recipient in place of the reproduction of the actor.  

All of the direct and indirect fitness changes are assigned to the actor. This assignment of the different pathways of genetic transmission to the actor associates all fitness changes with the behavior that caused those changes.  Hamilton viewed this inclusive assignment\break

\begin{figure}[H]
\begin{minipage}{\hsize}
\parindent=15pt
\noterule
{\bf \noindent\BoxLabel. Topics in the theory of natural selection}
\noterule
\noindent This article is part of a series on natural selection.  Although the theory of natural selection is simple, it remains endlessly contentious and difficult to apply.  My goal is to make more accessible the concepts that are so important, yet either mostly unknown or widely misunderstood.  I write in a nontechnical style, showing the key equations and results rather than providing full derivations or discussions of mathematical problems.  Boxes list technical issues and brief summaries of the literature.      
\noterule
\end{minipage}
\end{figure}
\boxlabel{preface}

\vskip -0.3in

\begin{figure}[H]
\begin{minipage}{\hsize}
\parindent=15pt
\noterule
{\bf \noindent\BoxLabel. Scope}
\noterule
\begin{quote}
Most research in biology is empirical, yet empirical studies rely fundamentally on theoretical work for generating testable predictions and interpreting observations. Despite this interdependence, many empirical studies build largely on other empirical studies with little direct reference to relevant theory, suggesting a failure of communication that may hinder scientific progress. $\ldots$ The density of equations in an article has a significant negative impact on citation rates, with papers receiving 28\% fewer citations overall for each additional equation per page in the main text. Long, equation-dense papers tend to be more frequently cited by other theoretical papers, but this increase is outweighed by a sharp drop in citations from nontheoretical papers \autocite{fawcett12heavy}.
\end{quote}

\noindent This article contains no equations beyond a few summary expressions.  I do not write for other theoreticians.  I do not attempt to be comprehensive.  Rather, I try to evoke some lines of thought that I believe will be helpful to scientists who want to know about the theory.  

A rough idea about the theory aids empirical study.  It also helps to cope with the onslaught of theoretical articles. Those theoretical articles often claim to shift the proper framing of fundamental issues. The literature never seems to come to a consensus.  

Here, I attempt to translate a few of the key points into nonmathematical summaries.  Such translation necessarily loses essential components of understanding.  Yet it seems worthwhile to express the main issues in way that can be understood by a wider audience.  Refer to my earlier work for mathematical aspects of the theory and for citations to the technical literature
\autocite{frank97multivariate,frank97the-price,frank98foundations,frank12natural,frank12naturalb,frank12naturalc,frank13natural}.      
\noterule
\end{minipage}
\end{figure}
\boxlabel{scope}

\noindent of all genetic consequences to the original causal behavior as the key to his approach.  The proper way of doing the calculation was the major theoretical advance.  \textcite{hamilton70selfish} emphasized this causal perspective:
\begin{quote}
Considerations of genetical kinship can give a statistical reassociation of the [fitness] effects with the individuals that cause them.
\end{quote}
Hamilton's 1970 paper also greatly advanced the analysis by using Price's equation \autocite{price70selection}.  With this method, Hamilton established the correct measure for genetic similarity, $r$, between actor and recipient. That measure turned out to be the regression coefficient of the recipient's genotype in relation to the actor's genotype.  That regression is the value needed to establish a proper measure of inclusive fitness. Hamilton noted that the regression was the exact measure only in his particular models.  He argued that the regression measure would extend, at least approximately, to a wide variety of other assumptions.  

Roughly speaking, $r$ translates the actor's deviation in gene frequency from the population average to the recipient's deviation in gene frequency from the population average.  Thus, recipient's reproduction has consequence for gene frequency change that is $r$ multiplied by what the same fitness increment in the actor would cause to gene frequency change \autocite[p.~49]{grafen85a-geometric,frank98foundations}.

\subsection*{Comparison of different approaches}

Already by 1975, different lines of thought on altruism had developed.  \textcite[pp.~336--337]{hamilton75innate} gave his opinion:

\begin{quote}
The usefulness of the `inclusive fitness' approach to social behaviour (i.e.\ an approach using criteria like $[rB-C>0]$) is that it is more general than the `group selection', `kin selection', or `reciprocal altruism' approaches and so provides an overview even where regression coefficients [$r$] and fitness effects [$B$ and $C$] are not easy to estimate or specify. 
\end{quote}
For Hamilton, his rule was a way to clarify biological understanding and to develop qualitative hypotheses about the causes of adaptation.  Hamilton did not think of his rule as replacing the way in which one did calculations for models of population genetics.  Indeed, he always considered the classical population genetics theory as the primary truth. He then evaluated his own methods in light of how well they could capture, in a simple way, the complexities of the underlying genetic models.  Following that historical line of thought, the subject subsequently split into two lineages.  

On the biological side, Hamilton's perspective completely changed the way people approach a great variety of key problems, ranging from social insects to parasite virulence to bacterial competition to the evolutionary history of ``individuals'' to the historical tendency for an increase in biological complexity.  Almost everyone agrees that this was a revolutionary change in biological thought and that it derived from Hamilton's work.  

At the same time, heated debates arose about theoretical interpretations and mathematical details.  We see in Hamilton's 1975 quote the different competing phrases and associated theoretical perspectives.  The ongoing debates have grown ever more fierce rather than settling out to a common perspective.  

\begin{figure}[H]
\begin{minipage}{\hsize}
\parindent=15pt
\noterule
{\bf \noindent\BoxLabel. Literature}
\noterule
\noindent For each topic related to kin selection theory, I list a small sample of key articles and reviews.  This limited space does not allow comprehensive coverage or commentary on the particular articles, but should provide an entry into the extensive literature and the range of opinions.

Several reviews follow Hamilton's perspective \autocite{alexander74the-evolution,dawkins79twelve,michod82the-theory,grafen85a-geometric,lehmann06the-evolution,wenseleers06modelling,dugatkin07inclusive,bourke11principles,gardner11the-genetical}.  An associated literature emphasizes the problem of sociality and sterile castes in insects, with additional commentary on general aspects of the theory \autocite{wilson71the-insect,west-eberhard75the-evolution,trivers76haplodiploidy,andersson84the-evolution,brockmann84the-evolution,alexander91the-evolution,bourke95social,queller98kin-selection,foster06kin-selection}.

Kin selection theory has been applied to a wide range of biological problems.  Here, I can list only a few general overviews. Those overviews give a sense of the scope but do not include many significant applications \autocite{trivers85social,maynard-smith95the-major,crespi01the-evolution,michod01cooperation,west07the-social,west09sex-allocation,burt08genes,davies12an-introduction}.

The strongest criticisms arose from population genetics.  The main issues concern how the specifics of genetics can vary from case to case and alter the outcome of selection, and how the full analysis of dynamics may provide an essential, deeper perspective on evolutionary process \autocite{uyenoyama81population,uyenoyama82population,karlin83kin-selection,kerr04what,nowak10the-evolution}.  

Kin selection theory has a long association with debates about units and levels of selection.  I give a very short listing, because that topic is beyond my scope \autocite{lewontin70the-units,dawkins82the-extended,keller99levels,okasha06evolution}. The related topic concerning group selection does fall within my scope \autocite{wade78a-critical,uyenoyama80theories,wade85soft,wilson83the-group,grafen84natural,nunney85group,heisler87a-method,queller92a-general,dugatkin94behavioral,soltis95can-group-functional,sober99unto,henrich04cultural,traulsen06evolution,west08social,leigh10the-group}.  

The merging of kin selection theory with quantitative genetics and multivariate analyses of selection follows various lines of development \autocite{cheverud84evolution,queller92quantitative,wolf98evolutionary,wolf99interacting,bijma08the-joint,mcglothlin10interacting,wolf10interacting}.  Advanced aspects of the theory and new directions of theoretical development continue to appear \autocite{rousset04genetic,grafen06optimization,taylor06direct,gardner07the-relation,fletcher09a-simple}.       
\noterule
\end{minipage}
\end{figure}
\boxlabel{literature}

In essence, I think there is almost no disagreement about how evolutionary process shapes biological characters.  No matter the perspective, when faced with the same biological problem, all of the different approaches usually arrive at roughly the same predictions about how evolution shapes characteristics.  Yet, in spite of that agreement, the arguments persist about whether one should call the underlying process `group selection' or `kin selection' or `inclusive fitness' or `population genetics' or whatever else is being promoted.  

Clearly we need to understand more than just the predicted outcomes: we must also understand the underlying causal processes.  So something is at stake here.  But what exactly?  The best way to understand that question is through the historical development of the subject.  So let us continue with Hamilton's 1975 article and subsequent work.

Almost everything that one would reasonably want to say about group selection in relation to inclusive fitness or kin selection is in the following quotes from \textcite{hamilton75innate}.  I quote in full because there has been much controversy and misunderstanding about these issues.  It helps to read Hamilton's perspective, given long before the current participants in the debates fully developed their views on the subject.
\begin{quote}
As against `group selection' it [inclusive fitness] provides a useful conceptual tool where no grouping is apparent---for example, it can deal with an ungrouped viscous population where, owing to restricted migration, an individual's normal neighbours and interactants tend to be his genetical kindred.
\end{quote}
In other words, inclusive fitness is more general. Group selection is just a case in which the positive association, $r$, arises from clearly defined aspects of groups. In cases for which groups are not easily delineated, the same underlying inclusive fitness approach still holds.  Continuing
\begin{quote}
Because of the way it was first explained, the approach using inclusive fitness has often been identified with `kin selection' and presented strictly as an alternative to `group selection' as a way of establishing altruistic social behaviour by natural selection \autocite{maynard-smith64group,lewontin70the-units}. But the foregoing discussion shows that kinship should be considered just one way of getting positive regression [$r$] of genotype in the recipient, and that it is this positive regression that is vitally necessary for altruism. Thus the inclusive-fitness concept is more general than `kin selection'. 
\end{quote}
Hamilton preferred to reserve `kin selection' for cases in which the positive regression, $r$, comes from interactions between individuals that we would commonly describe by terms of kinship, such as cousins.  I will later argue against Hamilton's use of `inclusive fitness' as the ultimate causal view.  I will end up using `kin selection' as the label for a wide variety of processes, because of the lack of a better alternative.  But in 1975, Hamilton's view made sense.  For now, it is useful to read what Hamilton said to understand his perspective on the different framings for causal process.  Continuing
\begin{quote}
Haldane's [\citeyear{haldane32the-causes}] suggestion about tribe-splitting can be seen in one light as a way of increasing intergroup variance and in another as a way of getting positive regression in the population as a whole by having the groups which happen to have most altruists divide most frequently. In this case the altruists are helping true relatives. But in the assortative-settling model it obviously makes no difference if altruists settle with altruists because they are related (perhaps never having parted from them) or because they recognize fellow altruists as such, or settle together because of some pleiotropic effect of the gene on habitat preference. If we insist that group selection is different from kin selection the term should be restricted to situations of assortation definitely not involving kin. But it seems on the whole preferable to retain a more flexible use of terms; to use group selection where groups are clearly in evidence and to qualify with mention of `kin' $\ldots$, `relatedness' or `low migration' (which is often the cause of relatedness in groups), or else `assortation', as appropriate. The term `kin selection' appeals most where pedigrees tend to be unbounded and interwoven, as is so often the case with humans.
\end{quote}
The point is that the different labels serve only to help identify the cause of association, $r$.  The underlying evolutionary process should be understood with respect to $rB-C>0$, even when it is difficult in practice to calculate directly the different terms in Hamilton's inequality. Hamilton favored analyzing complex biological problems with population genetic models, then interpreting those models in terms of the simple causal framework captured by the $r$, $B$ and $C$ components of his rule.

\section*{Using Hamilton's theory to solve problems}

\subsection*{Limitation of Hamilton's theory in practical applications}

\textcite{hamilton72altruism} used his rule to develop various qualitative hypotheses about social insect evolution.  When Hamilton analyzed kin interactions in quantitative models of sex ratios \autocite{hamilton67extraordinary} and dispersal \autocite{hamilton77dispersal}, he first made his mathematical calculations with genetic or game theory models, then made \textit{post hoc} interpretations of the quantitative results in terms of interactions between kin. Hamilton never used inclusive fitness theory or Hamilton's rule to solve for the quantitative phenotype favored by selection.  \textcite[p.~337]{hamilton75innate} noted that
\begin{quote}
Although correlation between interactants is necessary if altruism is to receive positive selection, it may well be that trying to find regression coefficients is not the best analytical approach to a particular model. Indeed, the problem of formulating them exactly for sexual models proves difficult (Chapter 2). One recent model that makes more frequent group extinction the penalty for selfishness (or lack of altruism) has achieved rigorous and striking conclusions without reference to regression or relatedness \autocite{eshel72on-the-neighbor}. But reassuringly the conclusions of both this and another similar model \autocite{levins70extinction} are of the general kind that consideration of regression leads us to expect. The regression is due to relatedness in these cases, but classified by approach these were the first working models of group selection.
\end{quote}

In 1979, I took Hamilton's graduate seminar at the University of Michigan.  I inherited Hamilton's interest in fig wasp sex ratios and the idea that one could develop models of kin interactions and sex ratios using the Price equation.  At that time, Hamilton was losing interest in working on such problems, and I was left with the last seeds of his insight on this subject.  As I pursued my empirical studies of fig wasp sex ratios, I also tried to learn how one could develop more realistic models of sex ratios with complex kin interactions.  

\subsection*{Using kin selection theory to analyze models of kin interactions}

At first, I had Hamilton's doubts about using kin selection theory directly to solve problems.  Instead, the method in those days was to solve the problem with population genetics, and then try to interpret the resulting predictions in terms of kin interactions.  The biological interactions from my empirical studies led to horrendously complex population genetic analyses \autocite{frank83theoretical}.  With great effort, I could solve some of the problems.  I repeatedly found that the \textit{post hoc} interpretations in terms of kin selection were simple and easy to understand.  For example, the value to a mother of an extra son is devalued by the mother's genetic relatedness to the competitors of her sons \autocite{frank85hierarchical,frank86hierarchical,frank86the-genetic}.  

This underlying simplicity in the exact results of population genetics led me to try Hamilton's suggestion about using the Price equation to model sex ratios.  Eventually, I found a simple Price equation method to get the same results as the complex genetic approach, when analyzing commonly used assumptions.  The Price equation method also gave results that were much more general than the population genetic methods. In a Price equation analysis, it was easy to follow the causal processes during the derivation, and it was easy to interpret the final results in terms of the biology.  The method was like reading sentences from a book in which the biological processes of competition, cooperation, and kin interactions were written in the clearest and most direct manner.  

The generality of my solutions, derived directly in terms of kin interactions, improved through a series of papers on sex ratios and dispersal \autocite{frank85are-mating,frank85hierarchical,frank86dispersal,frank86hierarchical,frank86the-genetic,frank87demography}.  At first, it did not make sense to twist the results for those biological applications into a form that looked like Hamilton's rule, $rB-C>0$.  The results did not appear in terms of simple costs and benefits.  Consequently, I gave little thought to Hamilton's rule, and instead thought only of how kin interactions shaped the evolution of interesting biological characters.  

\section*{Solving problems: dispersal example}

A model of dispersal illustrates how the application of kin selection theory changed through the 1970s and 1980s.  In the 1970s, \textcite{hamilton78evolution,hamilton79wingless} became interested in wing polymorphisms among insects.  For example, some of the parasitic wasps that live in figs have two male morphs. A wingless male stays within its natal fig to compete for nearby mates. A winged male leaves to search for mates in other figs.  Hamilton guessed that the dispersers must often die before finding another fig containing potential mates.  With such an extreme cost of dispersal, why would an individual develop to disperse rather than stay and try to mate locally?

\textcite{hamilton77dispersal} realized that competition between kin may explain dispersal even when the cost is very high.  A male that stays and outcompetes brothers replaces the brother's sperm with his genetically similar sperm, causing little gain in the net success of their shared genotype.  By contrast, any success of a male disperser against nonrelated males provides full benefit to the fitness gain of his genotype.  

Although \textcite{hamilton77dispersal} recognized the kin selection processes involved, they did not use kin selection theory as a method to analyze the problem.  Instead, they formulated a simple ecological model and then solved for the evolutionarily stable strategy (ESS). After obtaining the result, they then gave a \textit{post hoc} interpretation in terms of kin interactions.  

In 1986, I solved this problem in a much simpler and more general way by using kin selection theory as a method of analysis \autocite{frank86dispersal}.  This history of the kin selection approach sets the context for the modern understanding of the theory.  The following is a slightly modified summary from \textcite[Section 7.2]{frank98foundations}.

\subsection*{Hamilton \& May's ESS model}

\textcite{hamilton77dispersal} assumed that a habitat has a large number of discrete sites. In each year, the parents die after producing offspring. Each offspring has a trait that determines the probability, $d$, that it disperses from its natal patch. Those that stay at home, with probability $1 - d$, compete for one of $N$ available breeding sites. Dispersers die with probability $c$, and with probability $1 - c$ they find a patch in which to compete for breeding. All sites are occupied in the simple model discussed here. Hamilton \& May analyzed the case in which one breeding site $(N = 1)$ is available in each patch.  In an asexual model, Hamilton \& May found the ESS dispersal fraction, $d$, to be
\begin{equation}\label{eq:disperseAsex}
	d^* = \frac{1}{1+c},
\end{equation}
in which $c$, the cost of dispersal, varies from just above zero on up to nearly one.  Interestingly, even if the cost of dispersal is high and the chance of surviving the dispersal phase is low, nearly half of the offspring still disperse.  

In a second model, Hamilton \& May analyzed dispersal in a sexual species.  Sexual reproduction raised analytical difficulties, because some inbreeding is likely if mating takes place within the local patch.  Their method of analysis did not easily handle inbreeding, so they assumed that all males disperse before mating, and a fraction $d$ of females disperse.  With that setup, they found that the ESS dispersal fraction for females is 
\begin{equation}\label{eq:disperseSex}
	d^* = \frac{1-2c}{1-2c^2},
\end{equation}
in which the dispersal rate is zero when $c\ge1/2$.  Hamilton \& May understood the role of kin selection.  The dispersal rate is lower in the sexual compared with the asexual model because the genetic relatedness of competitors within the natal patch was less in the sexual model.  Less kin competition reduces the benefit of dispersing to avoid competition with kin, and so lowers the dispersal rate favored by selection.  Beyond that intuitive \textit{post hoc} interpretation, kin selection theory played no role in the analysis.  

\subsection*{Motro's population genetic analysis}

\textcite{motro82optimal,motro82optimalb,motro83optimal} analyzed a fully dynamic population genetic model for the same problem.  His complex analysis, spread over three articles, covered the same biological assumptions as Hamilton \& May, with a few minor extensions.  The length and complexity of Motro's work arose from the detailed population genetic analysis, as opposed Hamilton \& May's relatively simple ESS methods.  

Motro found that, in an asexual model, his result agreed with Hamilton \& May's expression given in \Eq{disperseAsex}. Motro obtained two additional results. First, in a sexual model in which the mother controls the dispersal trait of offspring, the same result arises as for the asexual model, as in \Eq{disperseAsex}. By contrast, when Motro tried to match the assumptions of Hamilton \& May for offspring control of phenotype, he obtained
\begin{equation}\label{eq:disperseOut}
	d^* = \frac{1-4c}{1-4c^2},
\end{equation}
which differs from Hamilton \& May's result in \Eq{disperseSex}. Motro drew two conclusions. First, in sexual models, the equilibrium depends on whether offspring phenotype is controlled by the mother or by the offspring. Second, under offspring control, explicit population genetic models failed to confirm Hamilton \& May's result. Motro attributed the mismatch to a failure of the simplified ESS method when compared with his exact and rigorous population genetic techniques.

\subsection*{Analysis by kin selection}

I analyzed this dispersal problem by kin selection in order to understand the different results of Hamilton \& May and Motro \autocite{frank86dispersal}. \Boxx{disperse} shows the expression for fitness and some technical details.  

Following \textcite{maynard-smith82evolution}, the standard ESS approach is to find a local maximum of fitness with respect to phenotype.  When the amount of genetic variation for the phenotype is small, that local maximum is an ESS.\break

\begin{figure}[H]
\begin{minipage}{\hsize}
\parindent=15pt
\noterule
{\bf \noindent\BoxLabel. Fitness expression for dispersal}
\noterule
\noindent In the dispersal model, fitness $w$ depends on the dispersal phenotypes at the three different scales: a focal individual, $d$; the average dispersal probability in the focal individual's patch, $d_p$; and the population average, $\bar{d}$, yielding
\begin{equation}\label{eq:disperseFit}
  w\left(d,d_p,\bar{d}\right) = (1-d)p\left(d_p\right) + d(1-c)p\left(\bar{d}\right).
\end{equation}
When an individual remains at home, with probability $1-d$, its expected success, $p\left(d_p\right)$, depends on the average dispersal fraction in its patch, $d_p$.  When an individual disperses, with probability $d$, its probability of landing in a new patch is $1-c$, and its expected success in the new patch, $p\left(\bar{d}\right)$, depends on the average dispersal probability in the population, $\bar{d}$.  The expected success expressions, $p\left(d_p\right)$ and $p\left(\bar{d}\right)$, can be written as
\begin{equation*}
  p(\Ga) = \frac{1}{1-\Ga+\bar{d}(1-c)},
\end{equation*}
in which one can use either $\Ga\equiv d_p$ or $\Ga\equiv\bar{d}$. The denominator is proportional to the number of competitors for breeding in a patch, and so the reciprocal is proportional to the expected success per individual in that patch.

When analyzing the fitness maximum with respect to phenotype, one takes the derivative of $w$ in \Eq{disperseFit} with respect to $d$.  That differentiation leads to terms in which one has the derivative of $d_p$ with respect $d$.  Such a term would require specifying how the average phenotype of neighbors in a natal patch, $d_p$, changes with respect to the phenotype of a focal individual, $d$.  With kin interactions, that relationship could be complex, because the focal individual's phenotype, $d$, would be correlated with the average phenotype of its neighbors, $d_p$, through the genetic similarity among neighbors.  

The lack of clarity about the slope of neighbor (or group) phenotype with respect to the focal individual's phenotype initially led to the abandonment of the simple maximization method for ESS analysis when genetic relatives interacted.  In 1986, when I first analyzed this dispersal problem, I published a Price equation method of analysis.  I also began to see that the simple ESS maximization method worked when one simply replaced the derivative of neighbor phenotype on actor phenotype by the coefficient of relatedness.  One could see the equivalence by matching up terms in a Price equation analysis with what one obtained by differentiating fitness with respect to phenotype and expanding out the terms.  However, in the 1980s, I was not certain about how to justify using the simple maximization approach, so I published only Price equation analyses.  Later, in 1996, with the help of Peter Taylor's deep understanding and elegant analysis, we published the general maximization method \autocite{taylor96how-to-make}.      
\noterule
\end{minipage}
\end{figure}
\boxlabel{disperse}

\noindent  At the time, it was generally thought that this ESS maximization method would not work with kin interactions, and so was not used to analyze problems of kin selection (\Boxx{disperse}).  

\noterulenote{\Boxx{disperse} near here}\medskip

I compared my Price equation analysis for the change in fitness with the terms obtained by an ESS maximization approach. The potentially difficult term in the ESS analysis is the slope (derivative) of average group phenotype with respect to individual phenotype. That slope from the ESS analysis always matched a regression coefficient in the Price equation.  Under common assumptions, that regression coefficient is exactly the coefficient of relatedness of an individual to its neighbors.  Thus, one can often replace the slope of group phenotype with respect to individual phenotype by the coefficient of relatedness from kin selection theory, $r$ (\Boxx{disperse}).

All of that may sound a bit complicated, but in practice it is quite easy.  Take the derivative of fitness with respect to the individual phenotype; set to $r$ the slope of average group phenotype with respect to individual phenotype; and solve for a local maximum to obtain
\begin{equation}\label{eq:disperseKin}
  d^* = \frac{r-c}{r-c^2},
\end{equation}
which is the result reported in \textcite{frank86dispersal} and discussed in more detail in \textcite[Section 7.2]{frank98foundations}.  

\subsection*{Kin selection simplifies and generalizes prior results}

Motro and Hamilton \& May assumed that only one female could breed in each patch.  They made that assumption because their methods did not allow them to analyze the more complicated situation in which multiple females bred in each patch.  To compare \Eq{disperseKin} to the prior models, let us first follow that earlier assumption of one breeding female per patch.  

If the organism is asexual, then in each patch the candidates for dispersal---the offspring of the single asexual mother---are related by a coefficient of one, $r=1$, and \Eq{disperseKin} reduces to Hamilton \& May's result in \Eq{disperseAsex}.  If the organism is sexual, and offspring phenotype is controlled by the mother, then $r = 1$, and we again have \Eq{disperseAsex}. The reason $r = 1$ with maternal control is that it is the relatedness of the individual that controls phenotype to the average phenotype of its patch that matters. With only one breeding female, the mother's phenotype and the average phenotype in the patch are the same, and $r=1$.

If, by contrast, offspring control their own phenotype, then relatedness among competitors is $r = 1/2$, because competitors on a patch are outbred full siblings. With $r=1/2$ in \Eq{disperseKin}, we recover the second result of Hamilton \& May in \Eq{disperseSex}. 

Hamilton \& May assumed that the mother mated only once, so that siblings are related by $r=1/2$. By contrast, Motro implicitly assumed that the mother mated several times and that offspring in a patch were only half sibs, so in his model $r = 1/4$.  Using that value of $r$ in the general solution of \Eq{disperseKin} yields Motro's result in \Eq{disperseOut}.  Thus, the single kin selection model of \Eq{disperseKin} explains the parent-offspring conflict and the difference between Motro's analysis and Hamilton \& May's model.

With the kin selection model, we are not limited to one breeding site per patch, or to an outbreeding system. Rather, we can treat $r$ as a parameter and express the ESS dispersal fraction in terms of the coefficient of relatedness. Higher relatedness increases dispersal. The reason is that a genotype competing with close relatives gains little by winning locally against its relatives. Even a small chance of successful migration and competition against nonrelatives can be favored.

\subsection*{Discussion of the new kin selection methods of analysis}

In retrospect, the kin selection analysis of dispersal seems simple and obvious.  Yet, at the time, Hamilton had not been able to use kin selection theory to analyze the problems of sex ratio and dispersal that interested him.  In the theoretical analysis of phenotypes, kin selection was only a \textit{post hoc} method of interpreting results obtained by other means.  The understanding of process was sufficiently confused that Motro could write three articles criticizing the Hamilton \& May model, arguing that only a formal population genetic analysis could give the correct results.  In fact, Motro simply made different assumptions about whether the dispersal phenotype was controlled by the parent or the offspring and about whether females mated once or many times.  Those distinctions become obvious from the simpler and more general perspective of the kin selection analysis of phenotype.  

The solution in \Eq{disperseKin} did not have any obvious connection to Hamilton's rule.  Similar kin selection analyses of sex ratio models also did not connect in any clear way to Hamilton's rule \autocite{frank85are-mating,frank85hierarchical,frank86hierarchical,frank86the-genetic,frank86dispersal,frank87demography}.  At that time, I had concluded that Hamilton's rule was not useful for solving realistic problems.  In application, nothing like Hamilton's rule appeared.  It turned out that I was wrong about the generality of Hamilton's rule, but it would take another ten years after 1986 to find the hidden connections.  Meanwhile, throughout the 1980s and early 1990s, debate about Hamilton's rule continued.  

\section*{Problems with Hamilton's rule before 1996}

Hamilton's rule became a widely used standard.  The simplicity of $rB-C>0$ allowed empiricists to think through how particular natural histories might influence the evolution of phenotypes and to formulate testable hypotheses---the essential attribute for a successful theory. Most biologists continue to abide by some notion along the lines of Hamilton's rule, based on its perceived success in explaining empirical patterns (citations in \Boxx{literature}).  

Yet many theoreticians vigorously attacked the simplicity of Hamilton's rule.  It appeared easy to set up scenarios in which the rule failed.  Those theoreticians who favored the rule replied with ever more sophisticated theoretical analyses.  Anyone with primarily biological interests, or lacking in years of specialized mathematical training, gave up following the details.  Clearly, the broader notions of kin selection uniquely explained diverse aspects of natural history.  Hamilton's rule seemed to capture the right idea, if not every possible assumption that one could conceive.  

In this section, I discuss some of the criticisms of Hamilton's rule.  I focus primarily on issues that arose before 1996.  In that year, my own understanding changed with the publication of \textcite{taylor96how-to-make}.  With the help of Peter Taylor's elegant insights, I came to see the proper generalization of Hamilton's rule.  That generalization united the simplicity of the original rule with a new and broader scope.  With the broader scope, what previously seemed like a long list of exceptions to the rule could be seen as part of an expanded way of framing problems of social evolution.  Later sections discuss the changes in understanding from 1996.  This section sets the necessary background by focusing on issues before that time.

\subsection*{Extra terms in Hamilton's rule}

Hamilton's rule is not sufficient if the direction of change favored by selection depends on some term in addition to $rB$ and $C$.  \textcite{queller85kinship} cited several earlier examples in which an extra term is required. He then showed the general way in which such terms arise.  Suppose that the phenotype, $x$, is the level of an altruistic behavior that is costly to an individual but beneficial to its neighbors.  Among all the neighbors that interact within a local patch, the average level of the altruistic phenotype is $x_p$.  Social interactions, through the effects of these behaviors, increment fitness by
\begin{equation*}
  w =  B x_p - C x,
\end{equation*}
in which $C$ is the cost to the individual for the behavior, $x$, and $B$ is the benefit received by the individual from neighbors that, on average, behave as $x_p$.  Putting that expression into the Price equation, one finds that the altruistic behavior increases when $rB - C > 0$, which illustrates Hamilton's rule.  Here, $r$ is the slope (regression) of $x_p$ with respect to $x$, which measures how strongly an individual's behavior is associated with the behavior of its neighbors.

Following a variety of earlier studies, \textcite{queller85kinship} pointed out that fitness might depend on synergistic interactions between an individual and its neighbors with regard to altruistic behaviors.  If the benefit only accrues when both the focal individual and its neighbor act in concert, then we need to consider a multiplicative term, $y=xx_p$.  One can think of this term as describing the phenotype of pairs of individuals, in which the phenotypic value of the pair depends on how each individual in the pair behaves.  For example, to achieve a task, it may be that both individuals have to contribute cooperatively to that task, otherwise no gain is achieved.  In many cases, one can think of the term $y$ as the average partnership phenotype of a focal individual paired with a randomly chosen partner from the group.  If we include this synergistic effect to the increment for fitness we have
\begin{equation}\label{eq:quellerD}
  w =  B x_p - C x + D y,
\end{equation}
in which $D$ is the fitness contribution of the partnership phenotype. Putting this expression in the Price equation yields the condition for the altruistic behavior to increase as $rB - C + \Gr D>0$. The term $\Gr$ is the slope of $y$ with respect to $x$, which measures the association between the individual's tendency to be altruistic and the tendency of that individual's partnerships to behave in concert when faced with a task that requires joint action.  

This analysis suggests that Hamilton's rule fails when there are multiplicative interactions between phenotypes, because factors beyond additive costs and benefits arise \autocite{queller85kinship}. To anticipate later discussion, note that we may think of $y$ in \Eq{quellerD} as any characteristic other than the focal individual's value for the particular behavior under study, $x$, and the average of that particular behavior in neighbors, $x_p$.  If $\Gr$ is the slope of $y$ on $x$ and $\Gr\ne0$, then the analysis will yield a condition for the increase of the altruistic character $x$ as $rB - C + \Gr D>0$.  It does not take much imagination to think of many different attributes, $y$, that could be associated with $x$ and therefore cause Hamilton's rule to fail, if one chose to think of the subject in this way.  Before developing that notion, let us continue with some additional issues.

\subsection*{Ecological context and density dependence}

\textcite{hamilton64the-genetical,hamilton64the-geneticalb} emphasized that limited migration would tend to keep genetically related individuals near each other. Such population viscosity could favor altruism through the increased relatedness of neighbors.   A popular series of papers in the 1990s raised a problem with Hamilton's view of population viscosity \autocite[Section 7.1]{wilson92can-altruism,taylor92altruism,taylor92inclusive,queller94genetic,frank98foundations}.  

In a viscous population, neighbors may be related and therefore candidates for altruism. However, those same neighbors may also be the primary competitors of a potential altruist.  Two potentially offsetting effects may occur.  First, altruism may increase the vigor and success of a neighbor, which provides a benefit to the actor in proportion to the relatedness between actor and recipient.  Second, the more vigorous neighbor may take more of the local resources, which imposes a cost on the original altruistic actor.  In some cases, the two factors may cancel each other.  If so, altruism cannot evolve in viscous populations, even though neighbors may be closely related.

\subsubsection*{An early analysis by Alexander}

The problem was expressed beautifully in a much earlier article that is rarely cited in this context \autocite[pp.~353, 376]{alexander74the-evolution}
\begin{quote}
Hamilton's development of the concept of inclusive fitness began with the argument that the reproductive success of an individual organism cannot be measured by alone considering the effects on the number and quality of direct descendants. Also involved are effects on the reproduction of genetic relatives. But, since both of these effects can only be measured in a comparative sense, there are always other individuals involved, and they are the reproductive competitors of the individuals and genetic elements being considered. In Hamilton's equations they [the competitors] are the population at large, an average of the rest of the species. Hamilton's arguments thus seem only to consider the detriments of altruism in terms of energy expenditure and risk-taking in the act itself, and to omit or at least not specify the problem of subsequent detriment to the altruist (or its descendants) owing to the presence of the recipient (or its descendants). But all of the members of the species, or population, will not compete equally directly with any given individual. Nearby individuals are more direct competitors. This would not affect Hamilton's calculations unless nearby individuals also have a greater likelihood of being closer genetic relatives. That such a correlation generally exists is obvious, and is acknowledged by \textcite{hamilton72altruism}. I believe this factor modifies every consideration of whether or not, and how, nepotism will actually evolve. $\ldots$ The significance of this problem can scarcely be exemplified better than by a point made earlier---that if degrees of relatedness and intensity of competition among individuals diminish together in certain, not unlikely fashions with distance from any given individual in a population, then nepotism cannot evolve.
\end{quote}

\subsubsection*{Dispersal and sex ratio}

Earlier, I discussed a model of dispersal by \textcite{hamilton77dispersal}.  In that model, selection favored dispersal to reduce competition against neighboring relatives and increase competition against distant, unrelated individuals.  That process of uncoupling the scales of relatedness and competition can be understood in light of Alexander's analysis.  

Another line of thought independently developed the relative scales of altruism, competition, and relatedness.   \textcite{clark78sex-ratio} argued that ``Competition between female kin for local limiting resources may explain a male-biased secondary sex ratio $\ldots$''  To a mother, the benefit of making an extra daughter is offset by the competitive effect that extra daughter may have on the mother's other daughters.  By contrast, sons may disperse before competing and therefore not reduce the fitness of other sons.

The development of sex ratio theory in the 1980s accounted for the interactions between dispersal, relatedness, the scale of competition, and the scale of resource limitation \autocite<reviewed in>{frank98foundations}.  That theory made clear that altruism and kin interactions could only be understood in the full ecological and demographic context of the behaviors under study.  As \textcite{alexander74the-evolution} emphasized, the scales of altruism and relatedness must be evaluated in relation to the scale of competition.  

\subsubsection*{Ecology and demography in relation to Hamilton's rule}

The terms $r$, $B$ and $C$ of Hamilton's rule depend on the ecological and demographic context.  Any consideration of natural history makes that clear.  Yet the subject is full of ``discoveries'' that Hamilton's rule fails because those terms are not constants, and that ecology and demography matter.  The tension arises because simple population genetic models tend to take costs and benefits as fixed parameters rather than ecologically derived variables that depend on context.  

Similarly, relatedness turns out to be part of a much broader problem of how to measure costs and benefits in common units.  The traditional view of relatedness translates gene frequency deviations in an actor with respect to gene frequency deviations in a recipient, putting all terms on the common scale of consequences for gene frequency change.  That makes sense.  However, much confusion arose because terms that appeared to be equivalent to relatedness often popped up in analyses, yet had a variety of meanings.  As we continue on through the history and the generalization of Hamilton's rule, we will see that Hamilton's rule can only be understood within a broader approach of partitioning the causes of fitness into meaningful components.  Before turning to that generalization, I continue with the criticisms of Hamilton's rule that dominated discussion in the 1980s.

\section*{Further problems: dynamics}

\begin{quote}
It is better to be vaguely right than exactly wrong \autocite{read09logic}.
\end{quote}

\noindent The basic principles of kin selection theory and its descendant ideas always hold.  Those principles are: costs and benefits of phenotypes matter; statistical associations between actors and recipients of behaviors matter; and heritability traced from the expression of phenotypes to representation among descendants matters.  To most biologists, kin selection theory is understood as a concise summary of those basic principles.  

The story differs in the theoretical literature.  Once one loses sight of the biology, all that is left concerns mathematical aspects of the theory.  Can a particular approach be used to make an exact calculation about predicted outcome?  If another approach is simpler but limited in the scope over which it is exactly right, should it be discarded entirely?

These questions primarily concern mathematical rather than biological issues. Why should you care about those questions if you are interested in the biology?  You should care because the theoretical literature has not done you the favor of sorting out the parts that matter to you versus the parts that do not.  Instead, each theoretical article seemingly replies to another theoretical article. The real conceptual progress that does matter remains buried under the weight of much that does not really matter to someone interested in biological problems.  

Ultimately, sorting all of this out can only be done at the technical level.  There is no way to argue that certain technical details do not matter without showing, technically, why they do not matter.  I expressed my views about the technical issues in my book \autocite{frank98foundations}.  Written 15 years ago, that book still gives a good overview of the issues.  Here, I continue to avoid technical discussion, and simply evoke my perspective on the main points.

\subsection*{Statics versus dynamics}

\begin{quote}
Often in the writings of economists the words ``dynamic'' and ``static'' are used as nothing more than synonyms for good and bad, realistic and unrealistic, simple and complex. We damn another man's theory by terming it static, and advertise our own by calling it dynamic. Examples of this are too plentiful to require citation \autocite{samuelson83foundations}.
\end{quote}

\noindent It would be helpful to calculate exactly how selection influences phenotypes.  What is the predicted sex ratio under certain patterns of mating and competition?  When will individuals band together to form cooperative groups? When will groups split apart or fail because of internal conflict?  The generalizations of Hamilton's rule and the broader theories of kin selection only provide exact calculations under certain assumptions \autocite{frank98foundations}.  

Roughly, the theory becomes exact when the variation in fitness is small. However, significant amounts of variation are nearly universal in biology. To use kin selection as an analytical tool, when can we assume that there is little variation?  Lack of variation primarily arises when a population approaches an equilibrium.  Near an equilibrium, little further change occurs, and the population comes nearly to rest---the condition of stasis. Analysis near an equilibrium is sometimes called `statics.'   As an exact analytical tool, kin selection theory is primarily a tool for statics.  

The problems of statics are widely known.  Real biological systems are unlikely to be near an equilibrium.  Thus, exact analysis must consider dynamics---the full processes of change.  Worse, many problems have different points at which the system could come to rest---alternative equilibria.  If one only analyzes what happens near an equilibrium, as in statics, one has no idea which of the alternative equilibria the system will end up near.  Only a full dynamical analysis of change can indicate which of the equilibria one would expect the system to evolve toward.

Given all of the benefits of dynamics and the limitations of statics, why would anyone ever consider a theory based on statics?  Because, to make a dynamic analysis, one has to make a lot of exact and very specific assumptions, otherwise one cannot do the analysis.  For example, one usually has to specify exactly how the genetics of a phenotype is controlled in order to make a complete model of population genetics.  The problem is that we do not know the proper assumptions to fill out the required list for a dynamic analysis.  So, in making all of the necessary detailed assumptions for a dynamic analysis, one is left with an exact calculation that applies exactly to nothing.  

By contrast, the static analysis requires few assumptions.  In kin selection arguments, usually one needs to specify how phenotypes and various environmental factors influence fitness.  One obtains a static analysis that translates the biological assumptions about fitness into a prediction about how natural selection influences the evolution of phenotypes.  Clearly, this is a vague sort of analysis, but it is not exactly wrong, as is the full dynamical analysis.  Put another way, a static analysis does not suffer the pretense of exactness. Instead, a static analysis accepts the limitations and calculates the qualitative predictions about what one expects to see in various natural settings.  

\subsection*{Comparative statics}

\begin{quote}
Now, an observer fresh from Mars might excusably think that the human mind, inspired by experience, would start analysis with the relatively concrete and then, as more subtle relations reveal themselves, proceed to the relatively abstract, that is to say, to start from dynamic relations and then proceed to the working out of the static ones. \textit{But this has not been so in any field of scientific endeavor whatsoever:} always static theory has historically preceded dynamic theory and the reasons for this seem to be as obvious as they are sound---static theory is much simpler to work out; its propositions are easier to prove; and it seems closer to (logical) essentials \autocite{schumpeter54history}.
\end{quote}

\noindent A static analysis summarizes the major forces that potentially influence phenotype.  However, that sort of simplified theory cannot predict the actual phenotypic value that one expects to observe.  Instead, one should think of statics in terms of comparison.  The dispersal model discussed earlier provides a good example.  In that model, the predicted dispersal probability from a static kin selection analysis was given in \Eq{disperseKin}, repeated here
\begin{equation*}
  d^* = \frac{r-c}{r-c^2},
\end{equation*}
in which $r$ is the relatedness between competitors on a patch, $c$ is the cost of dispersing, and $d^*$ is the predicted equilibrium dispersal rate.  Certainly, no one believes that this model will predict the actual dispersal rate in real cases.  Too many factors are left out.  Instead, the whole idea of the analysis is to isolate a few key processes.  

The relatedness term, $r$, raises another problem.  In an actual population, the relatedness in a patch would depend on the dispersal rate. The theory given above does not tell us how to analyze this dependence between relatedness and dispersal. To simplify the analysis, we could set relatedness, $r$, as a given parameter. The model then predicts that as the relatedness among competitors on a patch increases, the observed dispersal rate increases.  That sort of prediction is the method of comparative statics.  

Comparative statics begins with the assumption that a dynamic analysis following the joint dependence between dispersal and relatedness would, ideally, be preferable. However, one cannot easily achieve that ideal, because dynamical analysis requires specific assumptions about a variety of processes for which we do not have information.  Instead, one admits that one does not know enough to predict dynamics, and so the analysis should emphasize statics. A static analysis is based on fewer, simpler assumptions.

The comparative static analysis isolates key causal processes in a direct way.  For example, the dispersal model makes an interesting comparative prediction:  as the relatedness in a patch increases, the predicted dispersal rate increases.  The ideal empirical test identifies natural or laboratory settings in which relatedness changes and one can measure the associated change in the amount of dispersal.  The structure is: measure the change in the putative cause and the change in the outcome.  If the direction of change in the outcome repeatedly tends to follow the predicted direction of change, then one is on to something.  

\section*{A practical method for solving problems and recovery of Hamilton's rule}

Hamilton did not use kin selection theory to analyze models of phenotypes.  He did pass on to his students unpublished notes about how the Price equation might be used to study sex ratios.  I obtained those notes in 1979 while attending Hamilton's graduate seminar.  I modified Hamilton's Price equation method to solve a wide variety of problems with kin interactions, including the dispersal model that I presented earlier. Hamilton's class notes are posted as supporting information for this article in the files \textit{WDH\_notes\_part1\_SuppInfo.pdf} and \textit{WDH\_notes\_part2\_SuppInfo.pdf}.

The Price equation method was a bit tedious.  Eventually, I found the match between the standard ESS analysis of phenotypes and the Price equation analysis of kin interactions.  Earlier, I briefly mentioned the enhanced ESS approach for kin interactions.  This section describes that enhanced approach in more detail and considers the broader consequences for understanding the theory. 

\subsection*{First steps}

The Price equation is a general expression for the change in average phenotype.  To calculate the change in phenotype, one needs to specify the relation between phenotype and fitness.  For example, \Boxx{disperse} shows the relation between dispersal and fitness.  When one puts that expression for fitness into the Price equation, one obtains a variety of terms.  Each term describes the contribution of a component of fitness to the overall change in phenotype \autocite{frank86dispersal}.

In a typical problem with kin interactions, different components of fitness oppose each other.  Some favor the increase in the phenotype, others favor the decrease in the phenotype.  For example, dispersal imposes a direct cost on an individual, because the increased risk during dispersal increases the chance of death before reproduction.  By contrast, dispersal benefits the reproduction of neighbors by reducing the competition experienced by those neighbors.  Together, the various terms in the Price equation analysis define the different components of fitness. 

I found that I could avoid using the Price equation by directly calculating the value of the phenotype that maximized fitness.  The maximization approach came from analyses of ESS phenotypes \autocite{maynard-smith82evolution}.  The idea is simply that, at equilibrium with regard to selection, the favored phenotype must have fitness at least as great as any slightly different phenotype.  If that were not the case, then the nearby phenotypes with higher fitness would increase, and the previous candidate phenotype was in fact not an equilibrium with respect to nearby alternatives. 

To find a local maximum, one uses the standard calculus approach.  Write a function that relates phenotypes to fitness.  Take the derivative (change) of fitness with respect to phenotype.  That derivative describes whether fitness increases or decreases with respect to a change in phenotype.  One assumes that all members of the population have the same phenotype, and then analyzes the fitness change for a small fraction of the population that has a slightly deviant phenotype.  In other words, little variation exists.

A local maximum can only occur when the fitness neither increases nor decreases for the deviant phenotype.  If the derivative were increasing, then larger phenotypes would be favored.  If the derivative were decreasing, then smaller phenotypes would be favored.  When the derivative is zero, then a balance in forces has been achieved, and no change is favored.  When at a balance, one checks larger phenotypic deviations to make certain that fitness would indeed decrease if the phenotypic value changed.  If so, then the balance point, where the derivative is zero, is a local maximum and an ESS.  Maximization is just a mathematical trick for finding an equilibrium. An equilibrium is a point at which stasis occurs, leading to a static analysis.  

When applying the maximization method, one ends up with terms that are the change in partner phenotype with respect to the change in the focal individual's phenotype.  As I mentioned earlier, that kind of term was initially thought to be difficult to interpret when kin interactions occur.  When partners are genetically related, then the association between partner phenotype and focal individual phenotype depends on the degree of genetic similarity.  Early analysts recognized that relatedness matters \autocite{hamilton67extraordinary,hamilton77dispersal}, but abandoned the method because it was not clear exactly how genetic similarity would translate into the relation between partner phenotype and focal individual phenotype that arises in the calculus method of differentiation.  

I matched the Price equation terms to the ESS maximization method. I could see that the difficult term for the change in partner phenotype with respect to individual phenotype was like the regression term that \textcite{hamilton70selfish} had come to use as the coefficient of relatedness in his theory.  That equivalence meant that one could use the much simpler maximization trick and the power of calculus to analyze fitness and find ESS phenotypes.  One just had to replace the change in partner phenotype with respect to focal individual phenotype by the coefficient of relatedness  between them. 

Before 1995, I only published Price equation analyses.  That method, although a bit tedious, easily gave solutions to many problems of dispersal, sex ratio, and tragedy of the commons models for sociality.  I published a series of articles between 1985 and 1994 on those topics, as summarized in \textcite{frank98foundations}.  I did not publish the maximization method without use of the Price equation, because the Price equation had a prior history in the literature and seemed like a more defensible approach.  However, in \textcite{frank95mutual}, I analyzed a more complex problem, for which avoiding the Price equation and using only the calculus method provided important advantages in understanding the evolution of phenotypes in social interactions.  Thus, I needed to develop the calculus approach into a publishable form.

\subsection*{Generalized Hamilton's rule as a marginal value expression}

To develop the calculus method, I approached Peter Taylor in 1995.  Taylor found a way to connect the calculus method to Hamilton's rule.  I had abandoned Hamilton's rule, because nothing like Hamilton's rule had appeared in my numerous studies of different phenotypes.   Suddenly, all of my prior analyses could be understood more deeply by their connection to the generalized Hamilton's rule that came from our work \autocite{taylor96how-to-make}.  

The calculus method automatically separates out the various causes of fitness into the three aspects of Hamilton's rule.  First, all of the focal individual's phenotypic effects on its own fitness combine into one term.  In the calculus analysis, that term is the small change in direct fitness for a given small change in the focal individual's phenotype, holding constant the phenotype of other individuals.

The small changes are usually described as `marginal changes,' matching the classical usage of marginal values in economic analysis.  The first term is thus the marginal effect of an individual's phenotype on its own fitness, holding constant all other effects.  This marginal effect matches exactly the cost term in Hamilton's rule.  One traditionally defines this effect as a cost in such models, because in the standard case of altruism, one analyzes a phenotype that directly lowers the actor's fitness---the cost.  Mathematically, there is no need for this direct effect to be negative and costly, and in some models it is not.  But we retain the traditional usage and label this term `the cost.'  Because the calculus approach analyzes small changes, that method automatically gives us the marginal cost.  

The second component is the marginal effect of small changes in an individual's phenotype on the fitness of social partners.  Traditionally, the effect of the actor on recipients is called  the `benefit.'  Thus, this second effect is the marginal benefit component.

The third component measures the association between the actor's phenotype and the phenotype or genotype of social partners.  The exact measure depends on various issues \autocite{frank98foundations}.  Here, simply note that the calculus approach automatically weights any marginal benefit components by the association between the actor and recipient.  That association matches the coefficient of relatedness from Hamilton's rule, although in a generalized form.  

An equilibrium can occur only when selection does not favor a change in phenotype.  Thus, at equilibrium, the marginal Hamilton's rule equals zero and has the form
\begin{equation}\label{eq:mHR}
  rB_m - C_m = 0,
\end{equation}
in which I use the $m$ subscripts to emphasize that the benefit and cost terms are marginal values \autocite{taylor96how-to-make}.  The marginal values will change with changing phenotypic values and with changing ecological and demographic context.  Thus, this analysis makes clear that Hamilton's rule arises from context-dependent benefit and cost terms.  Others had noted the context dependence of those terms \autocite{grafen85a-geometric,queller92quantitative,queller92a-general}.  But \textcite{taylor96how-to-make} was the first approach that easily found the ESS phenotype and at the same time showed the underlying conceptual basis by expressing a generalized Hamilton's rule.  I use the term `generalized' because the analysis extended the kinds and complexity of social interaction that could be studied and the interpretation of relatedness.  It also became clear how to deal with multiple social processes simultaneously, leading to multiple marginal cost and benefit terms.

The Taylor \& Frank method leaves out much mathematical detail.  That simplicity allows one to start with an expression for how different phenotypes and other factors influence fitness, take a standard type of derivative from calculus, and evaluate an equilibrium ESS outcome favored by selection.  This method of analysis automatically gives the form for the equilibrium ESS condition as the marginal Hamilton's rule in \Eq{mHR}.  The ESS provides the basis for comparative statics, in which one can see clearly how the the predicted phenotype changes with respect to various social, ecological and demographic causes.  

To achieve that simplicity, one ignores dynamics, certain details of genetics, and the developmental complexities that connect genotype to phenotype. Most of those complications do not alter the mathematical results with respect to searching for equilibrium values.  That magical simplification arises because, whenever the amount of variation in the population is small, most of the complexities become negligible in size, and the method correctly identifies the outcome of selection.   Although it is possible to include additional complexity, the method simplifies by design, following the precepts of comparative statics.

\subsection*{Marginal Hamilton's rule analysis of dispersal}

The marginal Hamilton's rule in \Eq{mHR} evaluates a problem by the marginal costs and benefits of a phenotype.  Consider the dispersal model of \textcite{hamilton77dispersal} described earlier.  If we start with the fitness expression in \Boxx{disperse}, take the derivative of fitness with respect to the variant phenotype of a focal individual, and collect together the terms, we end up with the cost, benefit, and relatedness components of the marginal Hamilton's rule \autocite[Section 7.2]{frank98foundations}.

Following that method
\begin{equation}\label{eq:marginalC}
  C_m = \frac{c}{1-cd},
\end{equation}
in which the marginal cost of increased dispersal, $C_m$, is the cost of dispersal, $c$, divided by the level of competition on a patch for a breeding spot, $1-cd$ (\Boxx{marginalD}).  The marginal benefit is

\begin{figure}[H]
\begin{minipage}{\hsize}
\parindent=15pt
\noterule
{\bf \noindent\BoxLabel. Analysis of dispersal model}
\noterule
\noindent For the marginal cost in \Eq{marginalC}, we obtain the competition term in the denominator by counting the number of individuals on a patch competing for each breeding slot.  That number is proportional to the fraction of individuals that do not disperse and stay at home to compete, $1-d$, plus the fraction of individuals that disperse and become immigrants into each patch.  The immigrant fraction is the fraction that disperse, $d$, multiplied by the probability of surviving the dispersal phase, $1-c$.  Putting the pieces together for the denominator, which is an expression proportional to the number of competitors on a patch, we obtain $1-d + d(1-c) = 1-cd$.

For the marginal benefit in \Eq{marginalB}, the denominator measures the intensity of competition between pairs of individuals in the patch for a given level of dispersal and cost of dispersal.  Roughly speaking, intensity of competition can be measured by the probability that two individuals will compete for the same breeding spot.  From the previous paragraph, the intensity of competition for a breeding spot is proportional to $1-cd$.  Thus, the pairwise competition for a spot is proportional to $(1-cd)^2$.    
\noterule
\end{minipage}
\end{figure}
\boxlabel{marginalD}

\begin{equation}\label{eq:marginalB}
  B_m = \frac{1-d}{(1-cd)^2},
\end{equation}
in which the numerator, $1-d$, describes how increased dispersal reduces the competition experienced by neighbors.  The denominator adjusts the benefit of reduced competition by the intensity of competition between pairs of individuals for each breeding spot, $(1-cd)^2$ (\Boxx{marginalD}). Putting these terms in the marginal Hamilton's rule of \Eq{mHR} leads to the general solution for dispersal in \Eq{disperseKin}.  

The marginal cost and benefit expressions are essentially impossible to obtain either intuitively, by thinking about the dispersal problem, or by inspecting the mathematical expression for fitness given in \Boxx{disperse}.  It is only with the maximization technique that the separate marginal cost and benefit expressions can be found.  

Once one has the marginal expressions, one can study them to learn how the simple biological assumptions translate into cost and benefit effects on different components of fitness.  As the phenotype of interest changes, the costs and benefits change through complex social, demographic and ecological interactions.  There will essentially never be fixed costs and benefits that can be plugged into some equation. Instead, one must extract those costs and benefits from the biological assumptions and the analysis of the problem.  

Historically, people have tended to take Hamilton's rule as an expression based on fixed costs and benefits.  That history arose because of the population genetic modeling that was associated with the early evaluation of the theory.  In a population genetic model, the tendency has always been to set costs and benefits as parameters associated with different genotypes.  Frequency and density dependence, and other biological interactions, were considered to be distinct from the costs and and benefits, as if those costs and benefits were not an outcome of the biology.  That approach misled people to think of Hamilton's rule as an expression that translated fixed costs and benefits into a conclusion about evolutionary change.

The \textcite{taylor96how-to-make} method helps because it gives a way to translate biological assumptions about phenotypes and fitness into the separate marginal cost and benefit terms needed to evaluate Hamilton's rule.  One must keep in mind that Hamilton's rule is, in practice, an expression that derives particular meaning from context. Without context, the marginal cost and benefit terms are abstract.  That abstraction is the primary strength of Hamilton's rule, allowing it to apply universally.  At the same time, abstraction often causes confusion, because the power of abstraction requires a certain degree of consideration to understand and apply properly \autocite{frank12naturalb}.

\section*{Taylor's insight into reproductive value, demography and life history}

In the recent literature, the methods of \textcite{taylor96how-to-make} are primarily used to analyze specific models by the marginal version of Hamilton's rule (\Eq{mHR}).  As long as one accepts the approach of comparative statics, generating testable hypotheses follows a simple procedure. First, express the different phenotypes and ecological factors that affect fitness in an equation that describes the natural history assumptions.  Second, maximize the fitness expression to obtain the predictions of comparative statics.  

That mathematical method solves what had previously been complex and often beyond analysis. It also provided a conceptual advance by developing my earlier Price equation and maximization approaches into the marginal Hamilton's rule expression in \Eq{mHR}.  The marginal Hamilton's rule clarified understanding about how selection works and how to interpret a wide variety of problems in natural history \autocite{frank98foundations}.

\subsection*{The importance of reproductive value}

From my point of view, however, the great advance in \textcite{taylor96how-to-make} came from Peter Taylor's insight about reproductive value in models of kin selection.  Reproductive value concerns the relative contribution of an individual to the future of the population.  For example, older individuals may have a lower expectation of future reproduction before death than do younger individuals.  The benefit of altruism provided to an older individual must be discounted by the low expected reproductive value of old age when compared with the greater reproductive value associated with altruistic benefits given to a younger individual.

The problem of reproductive value had always been a part of kin selection theory.  \textcite{hamilton72altruism} clearly noted that different individuals in a social interaction may contribute differently to the future of the population.  In Hamilton's analysis, when calculating the benefit of altruism toward different individuals, one must weight each individual by its genetic relatedness to the actor and by its relative reproductive value.  For example, helping a young cousin in the prime of life provides greater net benefit than the same help given to a much older sibling near the end of life.  Relatedness alone is not sufficient.

\subsection*{Prior status of the theory}

Although the importance of reproductive value was understood, the theory was in an odd state before 1996.  The general approach used by Hamilton and most followers was simply to attach a reproductive value weighting to each component of fitness.  If an actor and recipient had different reproductive values, then an extended Hamilton's rule might be $rBv_r - Cv_a > 0$, in which $v_r$ and $v_a$ are the reproductive value weightings for recipient and actor.  If, for example, the recipient is a young individual near prime reproductive age, and the actor is an old individual near the end of life, then $v_r$ is much larger than $v_a$, and the old individual may be favored to express a costly altruistic act toward the young individual even if the recipient is only distantly related to the actor.  

Simply attaching reproductive value terms to fitness components is correct but often not helpful. It is not helpful, because it provides no guidance about how to find the right valuations in realistic problems.  

Separately, a complete theory of life history had developed \autocite{charlesworth94evolution}.  That theory provided clear guidelines for how to compare different classes of individuals and different components of fitness with respect to their reproductive values.  The relative reproductive value weightings predict how life history characters, such as relative investment in reproduction and survival, may change with age, condition, local ecological factors, and the demographic structure of the population.  That well developed theory of life history had many successes in explaining major aspects of organismal physiology and behavior.  

Actual problems of natural history often required combining the relative reproductive valuations with the role of kin interactions. However, connecting life history theory to kin selection analysis had not been achieved in a generally useful way.  When studying the theory of such problems, one could come up with various special approaches or complicated analyses \autocite<e.g., >{frank87demography}. However, if one cannot apply a theory in a clear and simple way, one does not truly understand the theory at a deep level.  In this case, the basic issue was combining reproductive value with the various aspects of phenotypic evolution that arose in social settings.  But it was not known how to do that in an easy way.

\subsection*{Combining kin selection and life history}

Peter Taylor saw all of that.  He figured out how to attach our general kin selection approach with life history analysis. The extended method accounted for the full context of ecological and demographic factors that interact with social phenotypes \autocite{taylor96how-to-make}.  The method easily translates natural history into analysis.  As before, one specified a problem by how phenotypes affect fitness.  In addition, one now also specified the different classes of individuals involved and the distinct components of fitness.  For example, there could be older individuals and younger individuals.  The age groups would be embedded in the demographic structure of birth and death rates, which could depend on the social phenotypes.  Each class could be either actor or recipient or both, in the sense that the phenotype of interest could be expressed by different kinds of individuals and could affect different kinds of individuals.  Many realistic problems of sociality have this sort of class structure.  

All of this may sound complicated.  But the beauty of Taylor's insight is that the crude maximization method that I had been using was transformed into a simple method that combined analysis of sociality, generalized notions of kin relations through correlations between phenotypes, and the full power of life history analysis.  

\subsection*{Opportunity for synthesis}

Up to 1996, I had relatively little interest in theories of kin selection, inclusive fitness, and group selection as separate subjects worthy of study.  Instead, I had thought of those theories as tools that one used to solve problems of natural history, in the sense of comparative statics.  I had developed those solutions into testable hypotheses about interesting phenotypes.  My collaboration with Peter Taylor opened my eyes to the deep problems of selection that had been latent in the subject.  

In particular, our new approach brought out the proper formulations for marginal valuation and reproductive valuation.  Of course, that was, in a sense, not new, because Hamilton had all the right ideas.  But, just as Hamilton could not use kin selection theory to solve the problems that most interested him, such as dispersal and sex ratio, he also could not use life history and reproductive value to solve problems of sociality embedded in the natural complexities of demography and multiple classes of social interactants.  

I had always believed that if one could not use a theory to solve problems, then one had only a superficial understanding of deeper concepts.  With \textcite{taylor96how-to-make}, I suddenly realized the deep connections between the understanding of marginal valuation and reproductive valuation in sociality and the solving of actual problems of natural history.  That new perspective caused me to take up the study of the general theory. I wanted to evaluate the status of the deeper principles in relation to the way in which one analyzed problems.  

\subsection*{Unresolved issues}

I soon found that \textcite{taylor96how-to-make} led to many key points that remained vague or unresolved.  Two issues stood out.  First, the meaning of the relatedness coefficient seemed to be confused. Variations in interpretation arose for different kinds of problems.  The original simplicity of Hamilton's genetic kinship theory had finally sunk.  There were many earlier hints of problems, but with the expanded scope of the theory, the contradictions became too numerous to ignore.  

The second issue concerned the connections between the models of phenotype in sociality and the developing theories of multivariate selection in quantitative genetics.  Those theories of multivariate selection had advanced greatly since the classic article by \textcite{lande83the-measurement}.  Following Lande \& Arnold, many studies considered how to partition the causes of selection among the different characters that affected fitness and how to analyze expanded notions of heritability.  Those advances in quantitative genetics seemed to be closely related to the problems of analyzing social evolution.  Some steps had been made to connect those advances to sociality, yet the deeper relations remained unclear.

Following up on what I learned by working with Peter Taylor, I set out to understand those two issues: the generalization of relatedness and the causal analysis of fitness and heritability.  I eventually came across a surprising number of problems that I had never understood or had not even realized that I was ignoring in my many prior analyses of kin interactions.

\section*{Queller's insight and the true meaning of Hamilton's rule}

In pursuit of unresolved issues, I soon came to the key articles by \textcite{queller92a-general,queller92quantitative}. Queller developed the idea that Hamilton's analysis had always been about the causal interpretation of fitness components.  Hamilton separated the total fitness effect of a social act into costs, benefits, and relatedness so that one could reason more clearly about how selection shaped behaviors.  Hamilton never presented the theory as an alternative to classical methods.  Instead, he was after causal decomposition.

Much of the literature through the 1980s lost sight of this primary emphasis on causal decomposition.  Controversies about population genetics versus kin selection were ultimately about the tension between dynamics and comparative statics.  Full dynamical analyses with detailed assumptions about genetics provide exact theories that perhaps apply to no real cases.  Statics applies approximately to all situations, but perhaps not exactly to any particular case.

In my own comparative statics analyses of dispersal, sex ratios, and various social traits, I focused on the solutions for phenotypes in terms of biological assumptions.  I did not try to analyze the problems with respect to the sort of causal decomposition into costs and benefits that Hamilton had emphasized.  Later I came to understand the power of the marginal Hamilton's rule. I then began to understand my past comparative statics models in terms of Hamilton's partition into causal components.  In particular, the marginal Hamilton's rule automatically separated fitness components into direct effects (costs) and indirect effects (benefits) weighted by relatedness \autocite{taylor96how-to-make}.  

\textcite{queller92a-general,queller92quantitative} had also derived an expression similar to the marginal Hamilton's rule.  Queller first partitioned fitness into components according to a regression model.  His regression terms included the phenotype of the actor and the phenotype of the recipient---or the genotypes depending on the analysis.  Queller considered the different phenotypes within the general context of quantitative genetic analyses of multivariate selection \autocite{lande83the-measurement}.  That connection to multivariate quantitative genetics eventually opened the way to a broader interpretation of the components of fitness and the components of heritability.

Queller put his multivariate regression expression for fitness into the Price equation to obtain a multiple regression form of Hamilton's rule.  One term is the effect of an actor's phenotype on its own fitness, holding constant the phenotype of its neighbors (cost).  The other term is the effect of the neighbors' phenotype on the actor's fitness, holding constant the actor's phenotype (benefit).  These cost and benefit terms arise as partial regression coefficients in the partitioning of fitness into components.  In the Price equation, one automatically obtains a weighting of the benefit term by the regression of neighbor phenotype on actor phenotype (or genotype).  That weighting provides a measure of relatedness.  Thus, the condition for the increase of an altruistic behavior is $rB-C>0$, where each term directly corresponds to a regression coefficient.

Queller's partial regression terms for cost and benefit are similar to the cost and benefit terms of the marginal Hamilton's rule.  However, two differences are important.  First, Queller's approach emphasizes the causal analysis of components.  He explicitly related the regression model of fitness to a path analysis model, which highlights a causal interpretation of the different terms in the regression.  

Second, the regression terms, although more general and potentially interpreted by causal analysis, provide a poor method for analysis of actual problems in terms of comparative statics.  Even simple models of dispersal and sex ratio lead to complex and essentially uninterpretable regression expressions.  That complexity led me to abandon the direct use of such regressions in my earlier Price equation models and instead to develop the maximization technique.  The maximization technique requires the additional assumption of limited variation.  In return for that assumption, one obtains a simple and powerful comparative statics tool.  In practice, one trades the conceptually powerful regression modeling and causal analysis for the analytically powerful maximization method and comparative statics.

Once I had the first formal expression of the maximization technique from \textcite{taylor96how-to-make}, I evaluated the advantages and disadvantages of the various approaches.  I could see that one had to unite Queller's causal approach through regression and path analysis with the analytical power of the maximization and marginal value techniques.  Causal analysis ties back to Hamilton's original goal of separating fitness and transmission into components in order to reason more clearly about how selection and genetics shape social traits.  Marginal values and comparative statics are also necessary, to provide the tools to analyze actual problems.  Put another way, causal analysis provides the foundations for reasoning about complex problems, and marginal value analysis provides the techniques for applying that reasoning to particular cases.  

\textcite{taylor96how-to-make} came close to uniting the quantitative genetic and causal models given by \textcite{queller92quantitative,queller92a-general} with the calculus techniques for comparative statics.  However, once that loose connection was made, I soon began to see the next level of unsolved problems.  In essence, the overly simple notions of an actor and a recipient and of a cost and benefit were too limiting.  Those restrictive assumptions limited general understanding of causes and the analysis of particular cases.  To move ahead, one had to think through various types of natural history and to work out how to separate the many causes into distinct components.  That causal decomposition was Hamilton's original goal.  But one no longer had to be confined to the oversimplified abstractions in which Hamilton originally worked to show how causal decomposition might be done.  Instead, the time had come for a more general approach.

The class structured modeling for social interactions introduced in \textcite{taylor96how-to-make} suggested the next steps.  Taylor \& Frank gave examples with multiple classes in social interactions.  Those sorts of multiclass problems lead to more general causal decompositions with multiple actors and recipients and, more generally, with a broader way to reason about the causal components of realistic problems.  Those multiclass problems also brought out the challenge of tracing the distinct pathways of transmission and heritability that determine what fraction of the change by selection transmits to the future population.

The way forward was to work carefully with causal decompositions of fitness and the transmission of characters, to use the Price equation as a formal tool to keep track of everything in an exact evolutionary analysis, and then to simplify as needed to obtain practical tools for comparative statics.  Although that may sound complicated, it turns out to be a natural extension of $rB - C > 0$.  One simply needs additional causal terms to match realistic problems, and to interpret the various terms more broadly than Hamilton did.  

\section*{The proper generalization of kin selection theory}

Hamilton's theory is ultimately about causal interpretation.  The proper generalization arises from a clearer understanding of causal decomposition.  With regard to the causal analysis of relatedness, \textcite{hamilton70selfish,hamilton75innate} had already given up on the limited interpretation of the theory with regard to pedigree relations and classic notions of kinship.  Instead, he realized that the theory could be extended to deal with genetic similarities no matter how those similarities arise.  Causally, selection must be indifferent to the process that generates genetic similarity. Selection can only act on the current patterns of genetic variation and the current processes that influence fitness.  

Subsequent generalizations continued to refine the causal interpretation of selection.  The theory naturally transformed from its initial emphasis on identity by descent and lineal kin relations to statistical associations of genotypes and then to broader aspects of correlated characters in social interactions.  

\subsection*{An example: interspecific altruism}

My study of altruism between species taught me that kin selection must be thought of as part of a wider set of problems \autocite{frank94genetics}.  I had asked the typical altruism question: when is an individual favored to help another at a cost to itself?  In this case, the problem concerned whether an individual of one species could be favored to help an individual of another species.  Clearly, the traditional view of genetic kinship could not be involved.  Members of different species that do not interbreed cannot be cousins or other types of related kin.  

I set up the interaction between species as a variant of Hamilton's standard model.  In this case, I evaluated whether altruistic behavior by one species toward a second species can increase by selection.

Individuals of the first species have phenotype $x$, the level of help they provide to individuals of the other species.  The altruistic phenotype directly reduces the fitness of actors from the first species by $Cx$.   Individuals of the other species have phenotype $y$, the level of help they provide to individuals of the first species. That altruistic phenotype of the second species directly enhances the fitness of recipients of the first species by $By$. The goal is to evaluate how these behavioral interactions determine the direction of change in the altruistic phenotype $x$ of the first species.

The first species cannot accrue inclusive fitness benefits by helping the second species.  An inclusive fitness benefit is the indirect transmission of the actor's genotype through the recipient of the actor's altruistic behavior.  Members of another species cannot carry genotypic differences that influence the evolution of traits in the focal species.  Inclusive fitness has no meaning in relation to altruism between species.  Nonetheless, we end up with Hamilton's rule, as follows.

Focus on an individual in the first species with altruistic phenotype, $x$.  That altruism reduces fitness by $Cx$.  The focal individual may receive benefits from the altruism of the other species.  Suppose that the particular social partners from the other species for the focal individual have phenotype $y$, which adds $By$ to the fitness of the focal individual.  The combination of gains and losses for these effects causes an increase in fitness to the focal individual when 
\begin{equation*}
  By-Cx > 0.  
\end{equation*}
Now comes the key step: the association between the altruistic behaviors of partners from the two species.  Suppose the slope (regression) of $y$ relative to $x$ is $r$.   Then, in evaluating fitness changes, we can use $y=rx$, because the altruistic level of the focal individual, $x$, predicts the associated value of $y$. The translation between $x$ and $y$ is the regression coefficient, $r$.  

Using $y=rx$, the fitness change is positive when 
\begin{equation*}
  Brx-Cx > 0, 
\end{equation*}
which is the same as
\begin{equation*}
  rB-C > 0.
\end{equation*}
Is that Hamilton's rule?  If one abides by inclusive fitness and the traditional view established by Hamilton, then the answer is no.  If one recognizes that the traditional view came into being before we understood the broader analysis of characters and the general role of correlations, then the answer is maybe.  In the latter case, we must figure out the broader context and its relations to traditional models of social evolution, and then make a decision about how to understand the full range of social characters.

After \textcite{taylor96how-to-make}, I followed up by trying to apply the new theory to various problems of natural history.   Several conceptual limitations became clear.  The most important problem concerned the meaning of relatedness.  A second associated issue concerned the interpretation of inclusive fitness.  The remainder of this section discusses relatedness.  The following section takes up inclusive fitness.

\subsection*{Two types of relatedness: social partners and transmission}

Queller's \citeyear{queller92quantitative,queller92a-general} quantitative genetic approach linked kin selection to Lande \& Arnold's \citeyear{lande83the-measurement} general analysis of multivariate selection.  In the traditional multivariate approach, one usually thinks of two different phenotypes as traits present in each individual.  That same conceptual approach to multivariate analysis can also evaluate social situations, in which one phenotype is present in one type of individual, and the other phenotype is present in the social partners of the first type \autocite{frank97the-price}.  

The two interacting types of individuals may be players in a game, members of different species, members of a family, or any other combination.  For any of those interactions, we can often evaluate the consequences in the same way that I described for the interaction between two different species.  The focal individual has a phenotype with direct cost $Cx$.  That focal individual also has social partners that influence the focal individual's fitness by a factor $By$, where $y$ is the average phenotype of the social partners.  These two phenotypes, $x$ and $y$, lead to a multivariate analysis of selection that depends on the correlation between characters.  In this case, the two characters happen to be in different individuals. But the analysis is essentially the same as a standard multivariate selection problem.  

The correlation between the focal individual's phenotype, $x$, and its partners' phenotype, $y$,  is best expressed as a regression coefficient $r$ of $y$ on $x$.  If partners are genetic kin, the $r$ will be a kind of genetic kinship coefficient.  If the partners are genetically similar but not by traditional lines of kinship, the $r$ is still similar to the form of the relatedness coefficient that \textcite{hamilton70selfish} introduced as the key to his theory, which did not depend on traditional kinship ties.  

The partners may have correlated phenotypes, but be genetically uncorrelated.  If so, our focal individual still gains by the beneficial social phenotype of its partners.  The magnitude of that gain is in proportion to the same regression coefficient, $r$, in which the association is purely phenotypic.  But we must separate two aspects: selection and transmission.  That separation forms a standard part of multivariate quantitative genetics.  Selection is the differential success within a period, such as a behavioral episode or a generation.  Transmission is the fidelity by which selected traits are transmitted to the future, the heritability.

In this model, we can think of two distinct classes of individuals.  The focal individual is both an actor and a recipient for its own phenotype, $x$.  It is an actor because it expresses the phenotype; it is a recipient because its success is influenced by the same phenotype, $x$.  The partners are actors, because they express the phenotype, $y$.  But they are not recipients, because we have not specified that either trait, $x$ or $y$, affects the partners' own success.  The focal individual is a recipient of the phenotype, $y$.  

The total fitness increment on the focal individual with respect to the phenotype, $x$, is proportional to $rB-C$, as shown in the previous section.  Evolutionary change depends on the heritability of the phenotype, $x$.  This follows the typical combination in genetics: selection determines relative reproduction, and heritability determines the fraction of selective change that is transmitted to the future.  

The heritability is not particularly important in this case.  Suppose, for example, that the heritability is $\Gt = Rh$.  Here, $R$ is the genetic kinship between the focal individual and its descendants, and $h$ is the fraction of the phenotypic variability attributed to genes.  In a typical parent-offspring example, $R=1/2$.  If we compared an individual to a niece through a full sibling, then typically $R=1/4$.  The condition for $x$ to increase must include the heritability.  Thus, the condition is 
\begin{equation*}
   \left(rB-C\right)\Gt>0,
\end{equation*} 
which describes the transmitted fraction, $\Gt=Rh$, of the selective gain, $rB-C$.  If the trait is heritable, $\Gt>0$, then this condition is the same as $rB-C>0$.  Here, distinguishing the similarity between social partners, $r$, and transmission in proportion to $R$, did not change the result.  

\subsection*{Distinguishing the types of relatedness}

In many  cases, one must distinguish the role of social partners from the role of transmission.  Consider the following example \autocite[Fig.~9]{frank97the-price}.  There are two classes of individuals.  Individuals of the active class express an altruistic phenotype, $x$, with cost $C$.  A focal individual of the active class has social partners from the active class that express an average level of altruism, $y$, with benefit $B$ to the focal individual.  The fitness increment for a focal individual in this active class with respect to the altruistic behaviors is $w_1=By - Cx$.  

Each individual from the active class also has a single partner from a second, inactive class.  Members of that second class do not express an altruistic phenotype.  They may, for example, be relatives that receive care but do not give care. The inactive partner gains a benefit, $\hat{B}$, from its active partner's altruism, $x$, so its fitness increment from altruism is $w_2 = \hat{B}x$.  To simplify, we ignore the potentially different reproductive values of the two classes. An inactive partner might, for example, be an offspring or a nondescendant kin.

From the previous section, the direct fitness effect on class one is $rB-C$, where $r$ is the regression of $y$ on $x$.  If the phenotypic association between an actor and its social partners, $r$, arises from genetic similarity, then $r$ is a classic kin selection coefficient of relatedness.  However, nothing in the model requires genetic similarity.  Here, $r$ only has to do with phenotypic similarity, because $rB-C$ is the fitness effect separated from aspects of transmission to the future.  

The class one individuals transmit their phenotypes to the future in proportion to $\Gt_1$, the heritability of their altruistic trait.  Thus, the total direct transmitted component by class one is proportional to $(rB-C)\Gt_1$, which matches the result of the prior section.  

In this case, we also have the beneficial effect, $\hat{B}$, of the class one individuals on their inactive partners from class two.  The benefit, $\hat{B}$, describes the fitness effect separate from aspects of transmission to the future.  For example, if the class two individuals are genetically unrelated to their actor partners, then the enhanced fitness of the class two individuals has no effect on the evolution of altruism, because those class two recipients are genetically unrelated to the actors.  

In general, we may specify the heritability of a class one actor's phenotype, $x$, through its beneficial effect on its class two partner, as $\Gt_2$.  For example, the class two partner may produce nieces and nephews of the actor, and $\Gt_2$ would be the relatedness of the actor to its nieces and nephews.  But the more general interpretation is important: $\Gt_2$ is the heritability, or transmitted information, of the class one phenotype through its beneficial fitness effect on its partner of class two.  The net direct effect of selection and transmission on class two is $\hat{B}\Gt_2$.  

Combining the class one and class two effects, the total transmitted consequence of the actor's phenotype favors an increase in altruism when
\begin{equation}\label{eq:directInc}
  (rB-C)\Gt_1 + \hat{B}\Gt_2 > 0.
\end{equation}
This expression combines the role of correlated phenotypes between social partners, $r$, and the pathways for the fidelity of transmission to the future, $\Gt$.  

This example illustrates how to combine the two different aspects of similarity, or relatedness, that arise in models of social evolution.  The general approach requires separating classes of individuals according to their role in the social process, following the direct fitness effects on each class, weighting each class by the fidelity of transmission of the phenotype under study, and also weighting each class by its reproductive value \autocite{frank97multivariate,frank97the-price,frank98foundations}.  In practice, one typically uses the maximization technique of \textcite{taylor96how-to-make}, as updated in \textcite{frank97multivariate,frank98foundations}.  

All of this follows the kind of causal decomposition at the heart of Hamilton's approach to kin selection theory.  But one has to accept several generalizations to the theory, otherwise the problem is beyond understanding by kin selection analysis.  In particular, one must separate the correlation between phenotypes that influences fitness from the correlation between an actor and various descendants that determines heritability.  

The original theories of kin selection and inclusive fitness blurred the distinction between these different kinds of correlation.  To understand the issues more broadly, one must accept a theory that follows different causes of fitness, including correlations between social partners, and different causes of transmission, including direct and indirect pathways by which phenotypes pass to future generations \autocite{frank97multivariate,frank97the-price,frank98foundations}.

\subsection*{Kin selection versus correlated selection}

I have emphasized a causal analysis of selection rather than a purely kinship analysis of selection.  In the broader causal perspective, the two key factors are transmission and selection.  Transmission of social characters always depends on aspects of shared genotype, or at least on shared heritable traits.  For selection, correlated social phenotypes play the key role.  Such correlations may arise by kinship, by shared genotype through processes other than kinship, or by associations through processes other than shared genotype.  

With regard to correlated social phenotypes, it may seem tempting to define the genetic associations as the proper limited domain for kin selection theory.  Hamilton developed his theory by first analyzing classical pedigree kinship. He then broadened his scope to include shared genotype through processes other than kinship.  But he did not expand his theory to the general analysis of correlated traits by processes other than shared genotype.

There can never be a final resolution with regard to the proper domain for kin selection theory.  Ultimately, subjective factors determine how different people choose to split domains and attach labels.  If someone chooses to associate genetic correlations with `kin selection' and nongenetic associations with `correlated selection,' that is fine as long as the choice is expressed clearly and understood as a subjective choice.

In the absence of analyzing particular problems, I would be inclined to separate kin selection and correlated selection.  Those processes seem different, and so it makes sense to differentiate between them.  However, I have repeatedly found that separating in that way is very unnatural when actually analyzing particular problems \autocite{frank98foundations}.  Neither the mathematics nor selection distinguish the way in which phenotypic correlations between social partners arise.  For example, in problems that follow the structure that led to \Eq{directInc}, the causal effect captured by the phenotypic correlation $r$ depends only on the phenotypic association.  The genetic aspects of transmission are handled independently by the $\Gt$ terms.

For the phenotypic associations, one could choose to separate the causes of association into shared genotype and other factors.  That separation would distinguish between a narrow interpretation of kin selection and a residual component of correlated selection.  That separation can certainly be useful.  But repeatedly, in analyzing particular problems and in developing the underlying abstract theory, the mathematics unambiguously leads one to blur the distinction when focusing on the causal analysis of how selection shapes phenotypes.  For the particular causal component that concerns differential success separated from transmission, it is only the phenotypic correlations that matter.

Intuition often runs against the lessons urged by logical and mathematical analyses.  That discord is perhaps the most interesting aspect of mathematics.  People tend to split over that discord.  Most trust their intuition above all else.  Some, having felt the failure of their intuition too many times in the face of unambiguous logic, give in to the mathematics. I follow the latter course, in which it is much better to adjust intuition to mathematics and logic than to try and bend mathematics and logic to fit intuition.  

In my view, the mathematics of selection has led inevitably to certain developments in the theory.  Over time, the theory came to subsume the early ideas of kin selection into a broader causal perspective. That broader perspective is much more powerful when trying to analyze particular problems, and much simpler and conceptually deeper when trying to grasp the fundamental principles of evolutionary change.  However, tastes vary. Others will prefer to separate and label differently.  If one properly understands the underlying theory, different labeling causes few problems and ultimately is not a particularly interesting issue.

\section*{Direct and inclusive fitness}

Consider two alternative ways to calculate fitness.  The direct fitness method counts only the direct reproduction of individuals.  If an individual behaves altruistically, we count only the negative effect of that behavior on the individual.  If that individual's social partners behave altruistically, then we add to the direct reproduction of our focal individual the benefit received from the altruism of neighbors.  To calculate the total effect over the whole population, we sum up all of the positive and negative effects on the direct fitness of each individual, based on the individual's own phenotype, the phenotype of each individual's social partners, and the heritabilities through different pathways of transmission.  

Inclusive fitness alters the assignment of fitness components.  If an individual behaves altruistically, we assign two components of fitness to that individual.  The negative cost of altruism reduces the individual's own reproduction.  The benefit of altruism to neighbors increases the neighbors' reproduction.  We assign that increase in neighbors' fitness to the original altruistic individual that caused that increase, rather than directly to the neighbors themselves.  We discount the neighbors' fitness component by their genetic similarity to the altruistic individual.  Thus, the individual that expressed the behavior is assigned both the direct effect  on its own fitness and the indirect effect on neighbors' fitness discounted by genetic similarity.

\subsection*{Hamilton's approach}

Hamilton's mathematical analysis showed that, under some conditions, inclusive fitness provides the same calculation as direct fitness.  Hamilton preferred inclusive fitness, because it assigns all fitness changes to the behavior that causes the changes.  Some of the fitness changes are direct effects on the individual expressing the behavior, and some of the fitness changes are indirect effects on other individuals receiving the behavior.  This assignment of all fitness effects back to the behavior that caused them provides a clearer sense of cause and effect.  Clear causal analysis aids in reasoning about the evolution of complex social behaviors.  For example, inclusive fitness emphasizes that the effects of a behavior on the reproduction of passive recipients can play a key role in determining whether genes associated with the behavior tend to increase in frequency.

Hamilton understood that direct fitness was the ultimate measure for evolutionary analysis.  His mathematical studies primarily had to do with showing that inclusive fitness was equivalent to direct fitness under many conditions.  Hamilton emphasized inclusive fitness as his primary contribution to understanding social evolution.  He discussed how inclusive fitness should be regarded as the fundamental process that encompasses kin selection, group selection, and other approaches to social interactions between genetically similar individuals \autocite{hamilton75innate}.  Almost all debates about the costs and benefits of Hamilton's approach and descendant ideas focus on inclusive fitness.

\subsection*{Development of the theory and failure of inclusive fitness}

Since Hamilton's initial work, the study of social evolution expanded to analyze a broader and more realistic range of evolutionary problems.  In my view, inclusive fitness has become as much a hindrance as an aid to understanding.  I am not saying that inclusive fitness is wrong. Inclusive fitness does provide significant insight into a wide variety of problems.   But one must know exactly its limitations, otherwise trouble is inevitable.  Realistic biological scenarios arise for which inclusive fitness is important but not sufficient.  When one does not clearly recognize the boundaries then, when faced with a solution for which inclusive fitness is not sufficient, it becomes too common to conclude that inclusive fitness and all broader approaches to kin selection analysis fail entirely, and one must discard the whole theory.  

The issues are somewhat technical in nature.  I provided a full analysis and discussion in \textcite{frank97multivariate,frank98foundations}.  Here, I give a sense of the problem and why it matters.  I begin by briefly summarizing the main points from the previous section, which distinguished alternative measures of association between individuals.  Separating those different kinds of association must be done clearly in order to understand the distinction between direct fitness and inclusive fitness.

In the previous analysis leading to \Eq{directInc}, two different classes of individuals interacted.  Consider first the direct fitness of class one individuals. They lose the cost $C$ for their altruistic behavior.  Their social partners from the same class express an altruistic behavior that provides an average benefit $rB$ to a member of the class.  The $B$ is the beneficial trait of partners per unit of costly phenotype expressed by each individual, and $r$ is the phenotypic association between the costly behavior of an individual and the beneficial phenotype of partners.  Thus, the total direct fitness effect on each individual of class one is proportional to $rB-C$.  The heritability of the altruistic phenotype for class one individuals is $\Gt_1$, thus the heritable increase in altruism from the direct effect of class one individuals is $(rB-C)\Gt_1$.

A second class of individuals does not express the altruistic phenotype, but may carry genes for that phenotype---for example, genetic relatives that receive care but do not give care. The net beneficial effect of altruism from class one on the direct fitness of class two individuals is $\hat{B}$.  The heritability for class two individuals of the altruistic trait expressed by class one individuals is $\Gt_2$.  Thus, the total heritable increase in altruism through the direct reproduction of class two individuals is $\hat{B}\Gt_2$.  Putting the direct fitnesses of the two classes together and weighting them equally leads to \Eq{directInc} from the previous section, repeated here for convenience
\begin{equation*}
  (rB-C)\Gt_1 + \hat{B}\Gt_2 > 0.
\end{equation*}

Note the two different kinds of association, $r$ and $\Gt$.  The $r$ coefficient measures the phenotypic association between the altruistic expression in social partners from class one.  It does not matter how that phenotypic association arises.  It may be caused by shared genotype, in which case it is a common type of genetic relatedness coefficient.  Or it may be caused by shared environment, such as sunlight or temperature, that is independent of genotype.  No matter the cause of the phenotypic association, the direct fitness of class one individuals is proportional to $rB-C$.  The actual value of $r$ is a regression coefficient, and is sometimes called a coefficient of relatedness.  However, it is more general than a coefficient of relatedness, because many different kinds of causes may be involved.  With regard to immediate evolutionary consequences, the cause of the association does not matter.  

By contrast, the $\Gt$ coefficients measure heritability, and so can reasonably be understood as a measure of genotypic contribution to the expression of the altruistic character.  In the case of $\Gt_1$, the measure is the heritability through the direct reproduction by an individual that expresses the altruistic behavior.  The $\Gt_2$ coefficient measures the direct contribution of class two individuals to the increase in the altruistic character, even though those individuals do not express the character.  Because $\hat{B}$ represents an increment in fitness caused by the behavior of class one individuals, the heritability of the altruistic phenotype expressed by class one individuals through the increment of fitness in class two individuals is proportional to the shared genotype between the class one actors and the class two recipients.  

If class two recipients are genetically unrelated to class one actors, then $\Gt_2=0$, and the condition for the increase in altruism is $rB-C>0$.  That has the form of Hamilton's rule.  However, $r$ measures phenotypic association, no matter the underlying cause.  It may be that social partners in class one are genetically unrelated but phenotypically associated.  Nonetheless, $rB-C>0$ is still the proper condition, although it is certainly not an inclusive fitness expression in the manner usually understood by that theory.  One can adjust definitions so that inclusive fitness still works.  But the clearest understanding comes from analyzing direct fitness, so that $r$ carries its natural interpretation as a phenotypic association that may be caused by shared genes or may be caused by some other shared nongenetic process.  

Alternatively, suppose that the phenotypic association between social partners in class one is zero, $r=0$.  Then the condition for the increase in altruism is 
\begin{equation*}
  -C\Gt_1 + \hat{B}\Gt_2 > 0. 
\end{equation*} 
Now consider the interpretation of the $\Gt$ coefficients in terms of transmission and heritability. The ratio of indirect heritability to direct heritability is $R=\Gt_2/\Gt_1$.  That coefficient, $R$, is the form of genetic relatedness commonly used in inclusive fitness theory.  For inclusive fitness, one measures the relative transmission of causal genes through indirect compared with direct pathways of reproduction, which is the ratio of heritabilities.  If, in the prior expression, we divide by $\Gt_1$, and use $R=\Gt_2/\Gt_1$, then we have
\begin{equation*}
  RB -C > 0,
\end{equation*} 
in which $R$ is the inclusive fitness coefficient of relatedness.  This form is the classic expression of Hamilton's rule, which we may interpret with respect to inclusive fitness.

The direct fitness approach gives the correct analysis in all cases, with proper interpretation of $r$ as a phenotypic association between social partners and $\Gt$ as transmission to the future through heritability.  Inclusive fitness arises as a special case.  By contrast, if one begins with an inclusive fitness perspective, one has to struggle to get the right interpretation, and confusion will often arise with regard to both the analysis and the interpretation.  

The actual distinctions between direct and inclusive fitness are more extensive and more subtle \autocite[Chapter 4]{frank98foundations}.  Direct fitness typically provides a clear and complete analysis, and subsumes inclusive fitness as a special case.  Inclusive fitness does have the benefit of an intuitively appealing causal perspective.  However, inclusive fitness is more limited and more likely to cause confusion.  As understanding of a subject develops, it is natural for yesterday's general understanding to become today's special case.  

\section*{Understanding how selection shapes phenotypes}

\textcite{hamilton70selfish} originally set out to develop a causal decomposition of social evolution into components.  His decomposition by inclusive fitness had two steps: the separation of fitness into components and the analysis of heritability.  With regard to fitness, Hamilton's approach partitioned the total effect of a phenotype into the direct consequence on the actor and the indirect consequence on social partners.  With regard to heritability, Hamilton weighted the different fitness components by their fidelity of transmission relative to the phenotype in the focal individual.  His coefficient of relatedness measured the ratio of the heritability through the indirect fitness component of social partners relative to the heritability through the direct reproduction by the actor.  

\subsection*{Multivariate selection and heritability}

Independently of Hamilton's work, the theory of natural selection developed during the 1980s and 1990s.  That development primarily followed the influential paper by \textcite{lande83the-measurement}, which built on two earlier lines of thought.  First, \textcite{pearson03mathematical} had established the partitioning of fitness into distinct components.  Second, \textcite{fisher18the-correlation} had established the modern principles of heritability and the conceptual foundations of quantitative genetics.  

The development of these two lines---the components of fitness and the components of heritability---independently paralleled the two lines in Hamilton's thought.  In retrospect, the parallel development is not surprising.  Anyone attempting a causal analysis of selection and evolutionary change would ultimately be led to those same two essential parts of the problem.  

In the early 1980s, I began  to develop my own methods for analyzing phenotypes influenced by kin selection.  I started with the Price equation methods that I inherited from Hamilton's graduate seminar in 1979.  As I refined my methods of analysis and then eventually generalized the approach with the help of Peter Taylor, I was inevitably up against the two problems of partitioning fitness into components and tracing pathways of heritability.  However, until my work with Taylor in 1996, I had not given much thought to the underlying structure of the problem.  

Following 1996, when I found Queller's papers that merged Hamilton's kin selection theory with the Lande and Arnold method of multivariate selection and quantitative genetics, I began work on developing that connection identified by Queller.  The result is that kin selection and inclusive fitness became part of the broader approach to the study of natural selection \autocite{frank97the-price,frank98foundations,frank12naturalb,frank12naturalc,frank13natural}.  With that advance, it is no longer possible to separate cleanly between the initial view of kin selection as a special kind of social problem among genetically similar individuals and the broader approach of causal analysis for phenotypes, fitness, and heritability.  

From the merging of kin selection theory and the broader aspects of selection and heritability, problems like the analysis of altruism between species have come to look like a kin selection analysis, and classical problems of kin selection have come to look like a Lande \& Arnold type of analysis of multivariate selection, with the addition of a more complex analysis of heritability.  

\subsection*{Statics and the three measures of value}

With regard to studying particular biological problems, I continue to favor comparative statics for its pragmatic approach.  In analyses of comparative statics, kin selection problems are transformed into the analysis of three measures of value: marginal value, reproductive value, and the valuations of relative transmission \autocite{frank98foundations}.  

Marginal values transform different phenotypic components into common units.  Suppose, for example, that we analyze the marginal costs of a behavior associated with the direct reproduction of an actor and the marginal benefits of that behavior associated with the indirect effect on the fitness of a social partner.  The relative marginal valuations provide a substitution, or translation, measure.  That measure tells us, for each small change in phenotype, how much the marginal benefits change relative to how much the marginal costs change.  For example, does a small change in the costs for direct reproduction translate into a small or a large change in the benefits for social partners?  It is the relative marginal valuations that give us that translation.  

Marginal valuation only applies to the analysis of small changes.  More generally, when one analyzes large changes, the regression coefficients from multivariate analysis arise.  In that context, the regression coefficients serve as translations for the relative scaling between different phenotypes and components of fitness \autocite{frank13natural}.  

Reproductive value provides a weighting for different kinds of individuals with respect to their contribution to the future of the population.  Reproductive value is a component of the transmission of phenotypes. However, we separate reproductive value from heritability, because reproductive value usually differs by demographic rather than genetic aspects.  For example, age is a demographic property, and individuals of different ages have different reproductive values, although they may have the same heritability in transmission of their phenotypes.  Ecological factors, such as available resources or the tendency for local extinctions of groups, also influence reproductive valuation.  In terms of classical demography, resource availability may affect birth rates, and local extinctions of groups may affect death rates.

Valuations of relative transmission, or heritability, obviously play an essential role in tracing the causes of evolutionary change by natural selection.  The coefficients of relatedness in the initial theories of kin selection had to do with relative heritabilities through different pathways of transmission.  Those coefficients of relatedness are just special cases of the broader analysis of heritabilities in the general study of natural selection.

Social evolution and traditional kin selection problems raise particular issues with regard to the three measures of value.  However, the analysis of those social aspects falls within the broader framework of natural selection, which applies to all problems of selection and the transmission of phenotypes to future generations.  This merging of kin selection and inclusive fitness into the broader framework for the study of selection has led to deeper understanding and more powerful analytical approaches.  At the same time, the separate initial history of kin selection compared with the analysis of nonsocial problems sometimes leads to confusion about the current understanding of the theory of social evolution.

\section*{The failure of group selection}

\begin{quote}
Properly understood, then, the origins of an idea can help to show what its real content is; what the degree of understanding was before the idea came along and how unity and clarity have been attained.  But to attain such understanding we must trace the actual course of discovery, not some course which we feel discovery should or could have taken, and we must see problems (if we can) as the men of the past saw them, not as we see them today.

In looking for the origin of communication theory one is apt to fall into an almost trackless morass.  I would gladly avoid this entirely but cannot, for others continually urge their readers to enter it.  I only hope that they will emerge unharmed with the help of the following grudgingly given guidance \autocite[pp.~20--21]{pierce80an-introduction}.
\end{quote}

\noindent A causal analysis of selection begins by expressing how phenotypes and other variables influence the fitness of individuals.  In social problems, the characteristics of an individual's local group sometimes enter the expression for individual fitness.  If group characters influence fitness, then a causal component of selection is attributed to the group.  That causal component attributed to the group is one common way in which group selection arises.  

Hamilton developed models of group selection by this partition into individual and group characters.  I used Hamilton's group selection methods in my own early studies.  Later, I came to understand the limitations and ultimate failure of the group selection perspective.  I then merged kin selection theory with the general causal analysis of selection and transmission.

\subsection*{Hamilton's group selection models}

In the 1970s, Hamilton studied social phenotypes in group structured populations.  He analyzed sex ratios and dispersal polymorphisms of wasps that live in figs \autocite{hamilton79wingless}.  Each fig formed a clearly defined group.  He also studied multigeneration groups of insects and other arthropods that lived in isolated rotting logs \autocite{hamilton78evolution}.  As always, Hamilton combined his natural history observations with mathematical models to analyze natural selection.  For these group structured problems, he followed the hierarchical multilevel methods of \textcite{price72extension}, as described in \textcite{hamilton75innate}.  

Interestingly, a Price equation analysis of group structured populations is similar to a Lande \& Arnold analysis of multivariate selection.  In the case of group structure, fitness depends on an individual's phenotype and on the average phenotype of social partners in the group.  That decomposition of fitness into individual and group components, when used in the Price equation, gives a causal decomposition that ascribes effects to the individual and group phenotypes.  The causal component attributed to groups may be interpreted as group selection. There is nothing special about using individual and group phenotypes in a Price equation analysis of fitness.  If one used an individual's  phenotype, the phenotype of the individual's mother, and temperature, one would get a decomposition in terms of those variables.  The analysis works for any choice of variables that affect fitness.

\textcite{lande83the-measurement} also used the Price equation for their analysis of multivariate selection.  Lande \& Arnold's approach was in fact very similar to the unpublished Price equation method Hamilton used to analyze the sex ratios of fig wasps.  However, Hamilton did not interpret his Price equation method broadly as the multivariate analysis of selection, but instead followed Price's limited interpretation of partitioning individual and group components of success.

Following Hamilton, I began my own studies of dispersal and sex ratios by thinking in terms of group structured natural history. The mathematical models partitioned individual and group components of fitness \autocite{price72extension,hamilton75innate}.   Hamilton was not a committed group selectionist in the sense that began to develop in the 1980s.  Instead, Hamilton interpreted group structure as one way to get a positive genetic association between individuals, as emphasized very clearly in the quotes from \textcite{hamilton75innate} that I presented in an earlier section.  To some extent, Hamilton's strong focus on group structure arose from his inability to analyze phenotypes such as dispersal and sex ratios in terms of kin selection and inclusive fitness.  He understood that those processes were the key, but he could not write down mathematical analyses in terms of kin selection.  He had access to Price's methods for group structuring and so used that method instead.  

Hamilton was not fully satisfied with his group level analysis of sex ratios as given in his 1979 notes from his graduate course at the University of Michigan.  He never published that analysis, perhaps because it showed only that greater genetic similarity within groups led to a stronger kin selection effect.  That vague point was already obvious, as he had emphasized in his 1975 article.  Simply to show that vague conclusion again for sex ratios did not add any real insight.  Hamilton's class notes are posted as supporting information for this article in the files \textit{WDH\_notes\_part1\_SuppInfo.pdf} and \textit{WDH\_notes\_part2\_SuppInfo.pdf} (see published version in \textit{Journal of Evolutionary Biology,} DOI given on first page of this article).

\subsection*{My early group selection models}

I took up the empirical study of fig wasp sex ratios in 1981.  At that time, I also began to study Hamilton's notes and to learn how to extend Price's hierarchical multilevel selection analysis to apply to my empirical work.  My initial success with that method was limited to seeing that greater group structuring and more limited migration increased the genetic similarity of individuals within groups. The more closely related group members are, the stronger the kin selection effects.  

By the reasoning from Hamilton's 1975 paper and his teaching, I understood the equivalence between the kin selection perspective and the conclusion that the greater genetic variance among groups, the stronger the tendency for biased sex ratios.  So one could call that a group selection perspective.  But there was always the clear understanding that the ultimate causal basis arose from kin selection or inclusive fitness perspectives.  I summarized my early understanding of hierarchical multilevel selection and related group selection analyses in my first article, \textit{A hierarchical view of sex ratio patterns} \autocite{frank83a-hierarchical}.

In my early work, I focused only on group structured populations. I maintained an unbiased dual perspective between kin and group selection.  However, in that early work, I made limited progress toward teasing apart how selection influenced sex ratio evolution. I was stuck at the same vague group selection perspective that stopped Hamilton.  I slowly figured out how to move ahead.  Particular sex ratio models played a key role---the synergism between application and abstraction.  

In a particular study, I analyzed the case in which males competed for mates locally within groups, females competed for resources against neighboring females, and the males and females migrated varying distances before mating.  I then traced the causal processes that determined the evolution of the sex ratio.  Assuming that the mother controlled the sex ratio of her progeny, one could adopt the mother's perspective with regard to pathways of causation.  This new work grew from Price and Hamilton's multilevel selection analyses, reflected in the title of a key article, \textit{Hierarchical selection theory and sex ratios.\ I.\ General solutions for structured populations} \autocite{frank86hierarchical}.  Although that article emphasized hierarchical multilevel selection, it also placed the group structured perspective into its proper role: a special case within the broader analysis of phenotypes by kin selection theory.

\subsection*{Pathways of causation replace group selection}

In the group structured sex ratio models, one can separate several distinct causes with respect to kin interactions.  For example, the value of an additional son depends on the mother's genetic relatedness to the males that her sons compete with for mates.   Greater relatedness reduces the transmission benefit to a mother for an additional son.  We can express the effect as the marginal gain in mating success through an additional son multiplied by the relative heritability of the mother's sex ratio trait through sons minus the marginal loss in mating success among competing males multiplied by the heritability of the mother's sex ratio trait through those competing males.  

The value of an additional daughter depends on the mother's genetic relatedness to the females that her daughters compete with for access to resources.  Greater relatedness reduces the transmission benefit to a mother for an additional daughter.  We can express the effect as the marginal gain in reproductive success for an additional daughter multiplied by the relative heritability of the mother's sex ratio trait through daughters minus the marginal loss in reproductive success among competing females multiplied by the heritability of the mother's sex ratio trait through those competing females.  In addition to the direct contributions through each sex, there are also effects of one sex on the other.  For example, an extra daughter may provide additional mating opportunities for sons.

The full analysis showed how various causal pathways influence the predicted sex ratio \autocite{frank85hierarchical,frank86the-genetic,frank86hierarchical}.  Those pathways often include the genetic associations between competitors, measured by coefficients of relatedness.  Typically, a coefficient of relatedness can be expressed either as the genetic variance between groups divided by the total genetic variance in the population, or as a regression coefficient that measures the genetic correlation between interacting individuals.  The two interpretations are simply alternative expressions for the same measure.  The first, group based expression for the measure suggests a group selection interpretation, whereas the second, individual based expression suggests a kin-centric interpretation.  However, the measure is the same in both cases \autocite{frank86hierarchical,frank98foundations}.  

The problem with the group-based interpretation is that different causal pathways may be associated with different patterns of grouping. Or there may not be any natural grouping.  The pairwise correlations of kin selection theory do not require group structure. If there is no group structure, kin selection works perfectly whereas group selection fails.

Group selection, which initially provided a nice intuitive way to think about group structured populations, ultimately proved to be the limitation in understanding the evolution of phenotypes.  Inevitably, I had to return to the fundamental causal level, in which the correlations between individual phenotypes and the different pathways of heritability were made clear.  

Once at the proper level of causation, one could see that emphasis on groups hindered analysis.  The proper view always derives from the causes of fitness and the pathways of transmission.  The different causes of fitness rarely follow along a single pattern of grouping.  For example, males may interact over one spatial scale, females may interact over another spatial scale, and mating between males and females may occur on a third scale.  It is relatively easy to trace cause by the correlations between phenotypes, the ecological context of resource distribution, and the pathways of genetic transmission.  Those causal components do not naturally follow the sort of rigid grouping needed for a group selection analysis to work.

The last paragraph of \textcite{frank86hierarchical} emphasized how analysis of particular biological problems led to a deeper understanding of causal process:
\begin{quote}
In summary, all three theories---inbreeding, within-sex competition among relatives, and group selection truly describe causal mechanisms of biased sex ratios in structured populations. Through the study of a variety of scenarios with hierarchical selection theory, I draw the following conclusions. First, inbreeding biases the sex ratio since producing a daughter that inbreeds $\ldots$ passes on twice as many parental genes as producing a son would. Second, as the amount of within-sex competition among related individuals increases, the relative genetic valuation of that sex decreases. Third, genetic differentiation among groups $\ldots$ and genetic correlation within groups $\ldots$ are related descriptions for the same phenomenon. Some recent papers \autocite{colwell81group,wilson81the-evolution} have stressed the group selection aspect of this phenomenon without clarifying its similarity to genetic relatedness. Using group selection for describing causal mechanisms is particularly slippery, since, as in the various scenarios presented in this paper, the differentiation among groups may refer to groups of competing males, groups of competing females, or groups that contain inbreeding pairs. While hierarchical selection theory, which is a group selection sort of analysis, has proved a powerful analytical tool, it seems that, for describing causal mechanisms, it is often useful to apply the genetic regressions [kin selection coefficients] considered in the discussion.
\end{quote}

\subsection*{Logically, there cannot be a group selection controversy}

Two conclusions emphasize the failure of group selection.  First, the ultimate causal processes concern correlations between phenotypes and pathways of genetic transmission.  Group structuring is just one limited way in which phenotypic correlations and genetic transmission pathways may be influenced.  Second, insisting on a group perspective greatly limits the practical application of the theory to natural history.  Most natural history problems do not have a single rigid group structure shared by all causal processes.  If one starts with a group selection perspective, solving problems becomes extremely difficult or impossible.  No gain in understanding offsets the loss in analysis.

Although group selection has problems that limit its scope, it also has attractive features.  There is a natural intuitive simplicity in group structured analysis.  Total selection arises from the balance between the dynamics of selection within groups and the dynamics of selection between groups.  Altruistic characters often tend to lose out during selection within groups and often tend to increase by selection between groups.  The problem is that once people gain such intuition, they do not easily give it up in the face of the inevitable conceptual and practical limitations.  Like the growth of any kind of understanding, one must allow the first general insights to become the special cases of broader conceptual and analytical approaches.  Pinning a topic to the first simple illustrative model limits progress.  Concepts and their associated language naturally develop and transform over time.

Given this history, the idea of a group selection controversy seems to me to be a logical absurdity.  I do understand the intuitive appeal of group selection.  I was trained by Hamilton to think about the interesting properties of groups and about the dynamical processes of within-group and between-group selection.  I also learned from Hamilton the mathematical techniques to analyze multilevel selection.  My early conversion, however, did not last.  Both the conceptual and practical limitations became apparent as I tried to make progress in understanding various problems of natural history and the mathematical models needed to evaluate those problems.  

In summary, when groups are the cause of genetic associations, then group structuring is the causal basis for the associations that drive kin selection.  When group structuring is less clear, the principles of kin selection still hold, as they must. 

\subsection*{However, the controversy continues}

The Los Angeles Times newspaper published an interview with E.\ O.\ Wilson on September 19, 2012.  With regard to kin and group selection, the interviewer began:
\begin{quote}
The biologist J.B.S. Haldane explained ``kin selection'' when he was asked whether he would lay down his life for his brother. No, he said, but he would for two brothers, or eight cousins. In the journal ``Nature'' in 2010 \autocite{nowak10the-evolution}, you challenged kin selection and created a stir, to say the least.
\end{quote}
Wilson replied (the following is an exact unmodified typographic transcription of the published article)
\begin{quote}
I was one of the main promoters of kin selection back when it looked good. By the `90s I thought I heard the whine of wheels spinning. Willie Hamilton's [\textit{British evolutionary biologist W.D. Hamilton}] generalized rule [of kin selection] was that if you've got enough people looking after [relatives], society could become very advanced. It wasn't working. By 2010 I had published peer-reviewed articles on what was thoroughly wrong [with kin selection]. I said we've got to go back to ``multilevel selection.'' Groups form, competing with one another for their share. It's paramount in human behavior. The spoils tend to go to groups that do things better---in business, development, war and so on. 

I knew the biology. I saw that multiple-level selection works, but in different ways in different cases, and [with] my mathematical colleagues, said [in the Nature article], kin selection cannot work. We knew that was going to be a paradigm changer. We published it and the storm broke.
\end{quote}
I agree with the three key points emphasized by Wilson in his article \autocite{nowak10the-evolution}: that multilevel selection is important; that one should think about group selection in human sociality; and that kin selection in relation to haplodiploidy is not sufficient to explain insect sociality.  However, I do have a different perspective on some of the conceptual issues and the history of the subject.

\subsubsection*{Multilevel selection}

I have emphasized Hamilton's own interest in multilevel selection.  Thus, Wilson's way of opposing kin selection versus multilevel selection does not make any sense to me.  To expand briefly on this point with regard to human sociality, note that Hamilton's \citeyear{hamilton75innate} primary publication on multilevel selection had the title \textit{Innate social aptitudes of man: an approach from evolutionary genetics}. In that article, Hamilton first developed his theoretical perspective on human evolution by extending Price's \citeyear{price72extension} hierarchical selection methods. Hamilton then devoted approximately ten pages to group level perspectives on human sociality.  

In Hamilton's \citeyear{hamilton96narrow} collected works, he gave the reprinted version of this article \autocite{hamilton75innate} the secondary title \textit{Friends, Romans, Groups}, and wrote in the preface for the article:
\begin{quote}
[I] am proud to have included the first presentation of Price's natural selection formalism as applied to group-level processes. $\ldots$ He [Price] himself published one application of the formula to groups but I think it was less explicit and general than mine, indeed almost as if he was trying still to conceal his formula's full significance \autocite{price72extension}.  For myself, I consider the format of analysis [for multilevel selection] I was able to achieve through his idea brilliantly illuminating.
\end{quote}

\subsubsection*{Kin versus group selection in human sociality}

\textcite{alexander79darwinism,alexander87the-biology} built his comprehensive evolutionary analysis of human sociality on the importance of group against group competition. In developing the theoretical foundations for his analysis, Alexander had thoroughly reviewed issues of group selection. He expressed his thinking on this topic in an article with the title \textit{Group selection, altruism, and the levels of organization of life} \autocite{alexander78group}.  

I took my first undergraduate course in evolution from Alexander at the University of Michigan in 1978.  His views on multilevel selection in \textcite{alexander78group} were particularly important in shaping how I understood the subject.  Interestingly, Alexander was not much influenced by Hamilton's multilevel selection analysis, which derived from the mathematical theories of Price.  Instead, Alexander was an entomologist with a deep interest in human behavior. He developed his thinking from broad consideration of natural history.

A comment on the first page of \textcite{alexander78group} provides historical context
\begin{quote}
[E]volution by differential extinction of groups has recently been modelled or discussed anew by several authors $\ldots$ E.~O.\ Wilson \citeyear{wilson75the-origin}, for example, has argued that ``In the past several years a real theory of interpopulation selection has begun to be forged, with both enriched premises and rigorous model building. $\ldots$ Insofar as the new theory considers the results of counteraction between group and individual selection, it will produce complex, nonobvious results that constitute testable alternatives to the hypothesis of individual selection. My own intuitive feeling is that interpopulation selection is important in special cases.'' 
\end{quote} 
Clearly, Wilson has long given thought to the potential importance of group structuring in nature.  The point here is that this line of thought goes back over 40 years, with much debate about conceptual issues and the relative importance for understanding natural history.  However, most theories conclude that differential extinction of groups is usually a relatively weak force \autocite{maynard-smith76group}.  Only Hamilton's milder version of group structuring in relation to differential reproduction and genetic differentiation between groups seems to be on solid ground as an explanation for common patterns of natural history.  

I have argued throughout that Hamilton's type of multilevel selection is clearly a special case within the broader theory of kin selection.  Although group extinctions are usually thought to be outside the scope of kin selection theory, the merging of demography and kin selection by \textcite{taylor96how-to-make} and \textcite{frank98foundations} brings those different processes within a single coherent theoretical framework.  Bringing together those different processes is more than just a theoretical convenience.  

Suppose, for example, that a particular problem requires analyzing how an altruistic phenotype evolves.  That phenotype may affect the probability of extinction for its group. Its group may be composed of genetically similar individuals, perhaps kin in the traditional sense.  The phenotype may also have other costs and benefits with regard to interacting with social partners.  It is too hard to figure out what to expect by separately analyzing extinctions, group variations in genetics, social processes of costs and benefits, and other processes.  \textcite{nowak10the-evolution} say that one must make a specific model for a specific case, and then one gets the right answer.  True, but the history of science shows unambiguously that one gains a lot by understanding the abstract causal principles that join different cases and different models within a common framework \autocite{frank12naturalb}.  That truth leaves only the sort of causal perspective I have emphasized as a reasonable candidate for how to think about such problems.

Certainly, not everyone agrees with me.  \textcite{alexander78group} understood Hamilton's \citeyear{hamilton75innate} claim that group selection was just a special case of kin selection.  However, they rejected that point of view
\begin{quote}
That groups are often composed of kin does not mean that kin selection and group selection are in any sense synonymous \autocite{west-eberhard75the-evolution,williams66adaptation,wilson75a-theory,wilson75sociobiology:}. As \textcite{west-eberhard76born:} points out, ``In the same trivial sense that kin selection is group selection, all of natural selection is group selection, since even `individual' selection really concerns the summed genetic contribution of a group---the individual's offspring.'' Moreover, although kin selection can occur in continuously distributed populations, group selection cannot. For reasons elaborated later, we agree with \textcite{maynard-smith71the-origin} that it is more appropriate to distinguish kin selection and group selection than to blur their differences by considering them together.
\end{quote}
This quote shows the clear historical precedent for Wilson's argument that multilevel selection is distinct from kin selection.  As often happens with historical analyses of ideas in science, one can find significant antecedents that support a variety of different positions.  Because a controversy about kin and group selection will always come down to how one chooses to interpret words, there can be no final resolution.  

What we can say, unambiguously, is that Hamilton never argued for a distinction between multilevel analysis and kin selection.  Instead, he saw multilevel analysis as one of the most powerful approaches to thinking about the general problems that arise in applications of his theory.  My own view follows Hamilton.  In addition, the mathematics do not allow a logical distinction.  Any distinction must be injected by a particular bias with respect to how one uses the words and interprets the history.  However, as long as the historical and conceptual issues are made clear, it does not matter to me how one chooses to use the words.  Indeed, given how clearly we understand the theory, it puzzles me why so much attention and argument continues to focus on this issue.

Returning to \textcite{alexander78group}, their main point concerned how we should think about human evolution.  In their concluding remarks about humans, they state
\begin{quote}
Human social groups represent an almost ideal model for potent selection at the group level \autocite{alexander71the-search,alexander74the-evolution,alexander75the-search,wilson73group}. First, the human species is composed of competing and essentially hostile groups that have not only behaved toward one another in the manner of different species but have been able quickly to develop enormous differences in reproductive and competitive ability because of cultural innovation and its cumulative effects. Second, human groups are uniquely able to plan and act as units, to look ahead, and to carry out purposely actions designed to sustain the group and improve its competitive position, whether through restricting disruptive behavior from within the group or through direct collective action against competing groups. 
\end{quote}
\textcite{alexander78group} certainly understood the broader implications of multilevel selection analysis with regard to a variety of biological problems.  The first paragraph of their summary is
\begin{quote}
[T]here may be few problems in biology more basic or vital than understanding the background and the potency of selection at different levels in the hierarchies of organization of living matter. The approaches currently being used by evolutionary ecologists and behaviorists in assessing the likelihood of effective selection at the level of groups or populations of individuals may also be used to advantage by those concerned with function at intragenomic levels. The kind of selectionist techniques used recently to analyze the behavior of nonhuman organisms may in the near future be widely applied toward understanding not only human social phenomena, but a variety of phenomena of classical biology such as mitosis, meiosis, sex determination, segregation distortion, linkage, cancer, immune reactions, and essentially all problems in gene function and in ontogeny.
\end{quote}
The year 1978 was a long time ago. I do not understand why these ideas are still thought of as novel or controversial.

\subsubsection*{Insect sociality}

Wilson particularly emphasized the failure of kin selection theory in explaining the evolution of advanced sociality in insects \autocite{nowak10the-evolution}.  Once again, we must consider what is meant by the scope of kin selection theory.  I tend to  think of kin selection as a particular causal perspective within the broader theory of natural selection.  Although it is tempting to limit the scope of kin selection theory to certain simple scenarios and predictions, such limits never make sense in the logical or mathematical analysis of the subject.  However, to the extent that others choose to distinguish more finely, that does not bother me as long as the concepts and history are made clear.  

For those who view kin selection narrowly, the potential problems of that narrow theoretical view for understanding the origin of complex sociality (eusociality) in insects was a lively topic in the 1970s and 1980s.  \textcite{andersson84the-evolution} introduces his excellent review of the topic by noting that eusociality has arisen multiple times in insects.  He then turns to an early theory based on kin selection to explain why eusociality is particularly common in bees, ants, and wasps (Hymenoptera).
\begin{quote}
Hamilton's \citeyear{hamilton64the-genetical,hamilton64the-geneticalb} celebrated explanation is that haplodiploid sex determination in Hymenoptera makes sisters share three quarters of their genes, whereas a daughter only receives half her genome from her mother. Hymenopteran females may therefore propagate their genes better by helping to raise reproductive sisters than by raising daughters of their own. Haplodiploidy therefore should make the evolution of nonreproductive female workers particularly likely among the Hymenoptera. This and other stimulating ideas of Hamilton's started a revolution in the study of social behavior, particularly of the role of kin selection \autocite{maynard-smith64group,michod82the-theory}.

Several entomologists have warned against overemphasis on the 3/4 relatedness hypothesis, and they have pointed to other factors important in the evolution of eusociality \autocite{kennedy66some,lin72evolution,alexander74the-evolution,michener74the-social,michener74were,west-eberhard75the-evolution,west-eberhard78polygyny,evans77extrinsic,crozier79genetics,eickwort81presocial,brockmann84the-evolution}. \textcite{hamilton64the-genetical,hamilton64the-geneticalb,hamilton72altruism} and \textcite{wilson71the-insect} also noted that haplodiploidy alone cannot explain eusociality in Hymenoptera. Such reservations were often forgotten, however, and the 3/4 hypothesis came to dominate many textbook and popular accounts. For example, in his comprehensive review of social behavior in animals, \textcite[p.~415]{wilson75sociobiology:} stated that ``the key to Hymenopteran success is haplodiploidy''  and that ``nothing but kin selection seems to explain the statistical dominance of eusociality by the Hymenoptera'' \autocite[p.~418]{wilson75sociobiology:}. A long list of similar evaluations of the 3/4 relatedness hypothesis by other authors could be cited.

The main empirical evidence in favor of the 3/4 hypothesis is that eusociality seems to have arisen many more times in the haplodiploid Hymenoptera than in other insects \autocite{wilson71the-insect,brockmann84the-evolution}. This evidence initially appeared impressive, but several recent findings indicate that haplodiploidy and 3/4 relatedness between sisters may have been of limited importance for the evolution of eusociality. Other factors have clearly been involved, and it seems possible that haplodiploidy has even been insignificant compared to these factors. At least five lines of evidence cast doubt on the overwhelming importance sometimes ascribed to haplodiploidy [and the narrowly defined kin selection hypothesis].
\end{quote}
The further details do not concern us here.  The main point is that by 1984, the problems with a narrow interpretation of kin selection for explaining eusociality had been widely discussed.

\subsubsection*{Summary}

Inevitably, the debates about kin selection and group selection will continue, because the ultimate problem concerns different usage of words.  People vary in whether they prefer to emphasize differences or similarities between components of a broader problem.  Those who like differences emphasize distinctions between kinship interactions and group structuring of populations.  Those who like similarities see kin and group selection as part of a broader theory of natural selection.  So what?  Perhaps the debate can advance a bit by a more nuanced consideration of the underlying concepts and history.

\section*{Discussion}

\subsection*{Separation into component causes}

Kin selection theory analyzes the evolutionary causes of social phenotypes.  Causal analysis is not an alternative to other analyses, such as population genetics.  Rather, causal analysis brings out the factors that one must emphasize to understand pattern.  Why do phenotypes vary in the way that they do?  What matters most?  What factors should one focus on to make testable predictions?

Hamilton separated evolutionary change into three causes.  A direct component affects the individual that expresses a phenotype---the cost.  An indirect component affects social partners influenced by the focal individual's phenotype---the benefit.  To combine those two components into the total evolutionary effect on the phenotype, one must adjust the indirect component to have the same units as the direct component.  The adjustment translates the indirect component of change into an equivalent amount of direct change.  That translation is often a genetic measure of similarity between the individuals affected directly and indirectly---the coefficient of relatedness.

Note the structure of causal analysis.  The total change is what matters.  To understand that total change, we separate it into parts that help to reason about the problem.  Once we have a set of distinct parts, we have to combine those parts back into a common measure for total change.  To combine properly, each part is weighted by a factor that translates into a consequence for total change.  Separation into distinct causal processes leads to testable predictions about which causal components explain patterns of variation.  Separation also highlights the common causal basis that unites previously unconnected problems within a common conceptual framework.

Causal analysis by kin selection is never an alternative to other analyses of total change.  Rather, it is a powerful complement to other approaches.  In practice, kin selection can be so powerful in analysis and so helpful in the conceptual framing of problems, that one does not need other complementary methods.  However, determining the best methods always depends on the particular goals.  For example, in the study of alternative genetic assumptions and complex aspects of dynamics, population genetic models provide superior methods.   

\subsection*{Hamilton's rule}

Confusion over Hamilton's rule arises when it is not properly understood as a partitioning of causes.  The rule is the partition of total change for a social phenotype into direct and indirect components.  It does not make sense to consider whether the rule is true or false.  Rather, following Hamilton, one thinks of the rule in two ways.  First, is there a simple form for the partition of causes that matches the ultimate measure of total change, at least approximately and under particular conditions?  If so, what are the proper definitions for the components?  Second, how should we expand Hamilton's original causal partition for more complex problems?  

Roughly speaking, Hamilton's original expression in terms of costs, benefits, and genetic relatedness provided a useful partition that works for simple problems.  However, as the theory was applied to more realistic problems, the associated causal analysis had to be extended.  The modern theory of kin selection provides a more comprehensive causal analysis.  Multiple direct and indirect components of fitness may occur.  Costs and benefits are understood to depend on context.  Relatedness coefficients and their generalization by multiple regression coefficients translate all fitness components into common units of total change.  Separation between selection and transmission clarifies the distinctions between causal components.  

Methods of analysis for solving problems have been developed to complement the causal decomposition.  The limitations of the analytical methods and the causal decompositions are reasonably well understood.  Causal decomposition and simplified analysis provide tools to enhance understanding rather than alternatives to more complex and detailed mathematical analyses of particular problems.

\subsection*{Limitations of inclusive fitness}

Hamilton introduced inclusive fitness as a particular type of causal partition.  Inclusive fitness assigns an indirect fitness effect through a social partner back to the behavior that caused the fitness effect.  For example, if an individual saves its sibling's life, that fitness benefit is attached to the individual who saved the life rather than the individual whose life was saved.  That fitness benefit is discounted by the genetic relatedness of the savior to the sibling.  

Inclusive fitness has the advantage of assigning changes in components of fitness to the phenotype that caused those changes.  That causal decomposition can provide much insight into evolutionary process.  The problem arises because that very particular form of causal partitioning is often equated with the entire theory of kin selection.  Instead, it is much better to view kin selection as a general approach to the causal analysis of social processes.  Inclusive fitness is a particular causal decomposition that helps in some cases and not in others.

For example, phenotypic associations between social partners that do not share a common genotype can have a very powerful effect on social evolution.  Inclusive fitness fails as a complete analysis of correlated phenotypes between social partners.  That failure does not mean we should give up on trying to understanding the causes of social evolution in such cases, or that we should conclude that kin selection theory fails as a general approach.  Instead, we must understand the broader approach of causal analysis, and how different aspects of natural history should be understood from a broader causal perspective.  That broader perspective was developed many years ago and has proved to be a powerful tool for analyzing complex social interactions.   

\subsection*{Correlated social phenotypes versus genetic relatedness}

When correlated social phenotypes do not arise from shared genotype, how should we think of the relatedness coefficients of kin selection theory?  Proper causal analysis solves the problem.  Changes in phenotypes cause changes in fitness.  Those fitness changes must be translated into changes in the transmission of phenotypes to the future population.  In analyzing the causes of fitness and the causes of transmission, we must put all the components together into a common measure of total change.  The weighting of the different components leads to different types of regression coefficients.  Those regression coefficients are the translations of different causal components into a common scale.  

The fact that Hamilton's original theory considered only a very particular aspect of transmission and genetic relatedness has led to confusion.  Hamilton's original regression coefficient of relatedness is not the single defining relatedness and regression coefficient of kin selection theory.  Rather, it is the particular coefficient that arises in the special inclusive fitness analysis that Hamilton considered in developing his theory.

\subsection*{Synergism between abstraction and application}

Abstraction arises by recognizing the common processes that recur in different cases.  Application demands analysis of particular phenotypes under particular circumstances.  Kin selection theory grew naturally by the synergism between abstraction and application.  Hamilton pulled out the first clear abstraction that united various simple applications.  Yet he could not use his abstract theory to move on to new applications.  In particular, he could not solve the problems of dispersal and sex ratios that arose from kin interactions.  

As the applied theory eventually developed for dispersal, sex ratios and more complex social phenotypes, deeper abstract principles emerged.  For example, the distinction between selection and transmission became clear, and relatedness coefficients became a part of translating causal components into common units.  The improvements in abstract theory enhanced the scope of application to complex social phenotypes.  

The necessary synergism between abstraction and application showed the ultimate failure of group selection.  In particular, group selection is a useful abstraction for a limited set of applications.  When faced with a variety of applications, such as sex ratio evolution with multiple male and female interactions, group selection fails.  Instead of the limited perspective of group selection, the deeper abstract principles dominate.  Those principles include a clear causal analysis of distinct fitness components, separation of selection and transmission, and the proper weighting of the distinct causal components to attain an overall analysis of total change.

When different people focus exclusively on either abstraction or application, deep tension and fruitless debate arise.  When the two modes come together, great progress follows.  

\section*{Acknowledgments}

National Science Foundation grant EF-0822399 supports my research.  

\bibliography{main}

%
%
%

\end{document}